\documentclass[aps,superscriptaddress,pra,preprint,floatfix,showpacs]{revtex4}
\usepackage{graphicx} \usepackage{amsmath,amssymb,latexsym}
\begin{document}

\title{Transient localization in the kicked Rydberg atom } \author{E.
  Persson} \affiliation{ Institute for Theoretical Physics, Vienna
  University of Technology, A-1040 Vienna, Austria, EU}

\author{S. F\"urthauer} \affiliation{ Institute for Theoretical
  Physics, Vienna University of Technology, A-1040 Vienna, Austria, EU}

\author{S. Wimberger} \affiliation{ CNR-INFM, Dipartimento di Fisica
  ``Enrico Fermi'', Largo Pontecorvo 3, I-56127 Pisa, Italy, EU}

\author{J. Burgd\"orfer} \affiliation{ Institute for Theoretical
  Physics, Vienna University of Technology, A-1040 Vienna, Austria, EU}

\date{\today}


\begin{abstract}
  We investigate the long-time limit of quantum localization of the
  kicked Rydberg atom. The kicked Rydberg atom is shown to possess in
  addition to the quantum localization time $\tau_L$ a second
  cross-over time $t_D$ where quantum dynamics diverges from classical
  dynamics towards increased instability.  The quantum localization is
  shown to vanish as either the strength of the kicks at fixed
  principal quantum number or the quantum number at fixed kick
  strength increases.  The survival probability as a function of
  frequency in the transient localization regime $\tau_L<t<t_D$ is
  characterized by highly irregular, fractal-like fluctuations.
\end{abstract}

\maketitle

\section{Introduction}
\label{sec:intro}
The quantum mechanics of classically chaotic few-degrees of freedom
systems has become an intensively studied in the field of ``quantum
chaos'' \cite{stock,cas87}.  One key feature is quantum localization,
i.e. the localization of the quantum wavefunction while the
corresponding classical distribution shows diffusion
\cite{stock,cas87,berry}.  In periodically driven systems, this effect
has primarily been studied in the kicked rotor, e.g.
\cite{cas79,fish,adachi,rotor}, and in the Rydberg atom in a
sinusoidal electric field, e.g. \cite{jensen,bay,koch,buch98,buch95,leo}.
For both systems, quantum localization is closely related to Anderson
localization in transport in disordered systems
\cite{cas87,buch98,fish}.  One signal of this analogy are strong
fluctuations in the quantum system for different observables when
varying some external parameter.  A third system experimentally and
theoretically studied is the periodically kicked Rydberg atom (e.g.
\cite{jones,frey,brom,burg89,hill,KleSch,bucks,
  noordam,yo00,reinhold04}), i.e. a hydrogen like atom prepared in a
high-lying state and subjected to a sequence of ultra-short impulses.
In a recent publication, we showed the existence of quantum
localization in the positively kicked Rydberg atom \cite{eper03}, i.e.
the hydrogenic system with the initial state prepared in a highly
elongated quasi-one dimensional state localized on one side of the
nucleus and the periodic impulsive momentum transfer $\Delta p > 0$
pushing the electron away from the nucleus.  The classical phase space
of this system is globally chaotic with all tori destroyed for
arbitrarily small $\Delta p$, and the classical survival probability
decays algebraically, $P_{\rm sur}\propto t^{-\alpha}$, with
$\alpha\approx 1.5$ (see Fig.~\ref{fig:psurK} and \cite{hill}). In
contrast, by following the time-dependent system up to a few thousands
of kicks we could show both the quantum suppression of classical
ionization and the ``freezing out'' of the wavefunction, the hallmarks
of quantum localization.  Several issues remained open, however. They
include 1) The origin of a slow, yet noticeable, decay of the
localized states, 2) the dynamical role of high harmonics (up to
infinity) present in the system, and 3) the properties of strong
fluctuations present in the localization regime.

In the following paper we address these issues. We identify two
characteristic time-scales (``break times'') in this system. In
addition to the localization time $\tau_l$, where quantum and
classical dynamics begin to differ from each other due to universal
destructive interferences, there is a second break time $\tau_D$ where
localization is broken. Beyond $\tau_D$, a second cross-over occurs
where the classical dynamics becomes more stable. This second
cross-over in the kicked atom is related to the presence of
non-classical photoionization.

In the next section, Sect.~\ref{sec:meth}, we describe the method used
in our studies and in Sect.~\ref{sec:psur} we show that quantum
localization in the kicked Rydberg atom is transient. Strong
fluctuations in the survival probability as a function of frequency
for fixed times are studied in Sect.~\ref{sec:fluctuations} and in the
last section (Sect.~\ref{sec:summary}) a summary is given.

\section{Method}
\label{sec:meth}

The Hamiltonian of the one-dimensional kicked Rydberg atom is (in
atomic units)
\begin{equation}
H(t)=H_0-q\Delta p\sum_{k=1}^K \delta(t+T/2-kT) \;,
\label{eq:hamdelta}
\end{equation}
where $H_0=\frac{p^2}{2}-\frac{1}{q}$ is the hydrogen Hamiltonian and
$q$ the position of the electron. $\Delta p$ and $T$ are the strength
and the period of the train of kicks, respectively.  We will use the
number $K$ of kicks and time $t=K\times T$ interchangeably.  
The restriction to a 1D model in the present context is necessary
since the study of the long-time limit in 3D is currently 
computationally not feasible. Our previous studies for up to
$10^3$ kicks have shown that the 1D model can reproduce essential 
features of the 3D problem.
For a more detailed discussion of the relation between the 1D model 
and the real 3D dynamics, see \cite{eper03} and references therein.  
In the case $\Delta p>0$, i.e.
the kicks directed away from the nucleus, the classical phase space is
void of stable islands and the effect of quantum localization has been
shown to set in within a few hundreds of kicks \cite{eper03}.  In the
opposite case $\Delta p<0$ stable islands persist allowing for
survival of Rydberg states both classically and quantum mechanically,
referred to as stabilization \cite{yo00}.

The unidirectional kicks build up an average field $F^{\rm av}=-\Delta
p/T$. Hence, the time-periodic Hamiltonian (\ref{eq:hamdelta}) can be
decomposed into the time-independent Stark Hamiltonian
\begin{equation}
H_{\rm Stark}=H_0+qF^{\rm av}
\label{eq:hamstark}
\end{equation}
plus an infinite series of harmonics of equal strength,
\begin{eqnarray}
H(t)=H_{\rm Stark}+2F^{\rm av}q\sum\limits_{m=1}^{\infty}
\cos\left(2\pi m/T\left(t-\frac{T}{2}\right)\;\right) \; .
\label{eq:hamfourier}
\end{eqnarray}
In this paper, we will highlight the influence of the higher harmonics
$m>1$, their presence distinguishing our system from the Rydberg atom
driven by a microwave field.  For the ``positively kicked'' Rydberg
atom with $\Delta p>0$ (i.e. $F^{\rm av}<0$) the Stark Hamiltonian
possesses a potential barrier with maximum at $E^{\rm
  barrier}=-\sqrt{2|F^{\rm av}|}$ and $q^{\rm barrier}=\sqrt{1/|F^{\rm
    av}|}$ resulting in a finite number of quasi-bound states and a
continuum.  To keep the Stark Hamiltonian invariant, i.e. the average
field fixed, we vary $\nu$ and $\Delta p$ such as to keeping the
average field $F^{\rm av}$ fixed.

To calculate the long-time evolution of the quantum system we
represent the period-one time-evolution operator
\begin{equation}
U(T)=\exp(-iH_0\;T/2)\,\exp(i\Delta p \,q)\,\exp(-iH_0\;T/2)
\label{eq:floqop}
\end{equation}
in a basis $\mid n \rangle$ defined by $H_0\mid n \rangle = E_n \mid n
\rangle$ by means of the pseudo-spectral method \cite{tong}.
Dirichlet boundary conditions are applied at $q=0$ and $q=q_{\rm
  max}$.  Solving the eigenvalue equation $U(T) \mid
\phi_j\rangle=\exp(-iT{\cal E}_j) \mid \phi_j\rangle$ yields the
time-dependent wavefunction in terms of Floquet states $\mid \phi_j
\rangle$ as
\begin{equation}
\mid\psi(KT)\rangle=U(T)^K\mid\psi(0)\rangle=
\sum\limits_j d_j \exp(-iKT{\cal E}_j)\mid\phi_j \rangle
\label{eq:psiT}
\end{equation}
with $d_j=\langle \phi_j \mid \psi(0) \rangle$.  A masking function in
$q$ is introduced to avoid spurious reflections at $q_{\rm max}$ (see
\cite{eper03}).  For the low frequencies used in this paper, we apply
the masking operator three times per period.  The convergence of the
wavefunction obtained is tested both by varying $q_{\rm max}$ and by
comparison with direct solutions of the time-dependent Schr\"odinger
equation (i. e. without facilitating the Floquet states $\phi_j$) as
described in \cite{eper03,tong}.

For later reference we introduce scaled units, denoted by the
subscript $0$.  They leave the classical dynamics invariant and are
defined by $E_0=E\times n_i^2$, $T_0=T/(2\pi n_i^3)$, $\nu_0=1/T_0$,
and $F_0^{\rm av}=F^{\rm av}\times n_i^4$ where $n_i$ is the principal
quantum number (action) of the initial state \cite{koch}.  Note that
the transition energy due to to absorption of one photon
$E_0^\gamma=n_i^2\times \hbar\omega= n_i^2\times 2\pi/T= \nu_0/n_i$,
is {\it not} scaling invariant.

\section{Transient quantum localization}
\label{sec:psur}

\subsection{Survival probability and effective quantum number}

One measure to study quantum localization is the survival probability,
defined as $P_{\rm sur}(K)=\langle\psi(KT) | P_{\rm bound} |
\psi(KT)\rangle$ where $P_{\rm bound}$ is the projection operator onto
bound hydrogenic states.  The underlying picture is that transport
along the energy axis towards the continuum is considered to be the
equivalent of conductance in disordered systems \cite{borgonovi}.
Correspondingly, suppression of energy absorption from the pulse
sequence and, thus, suppression of ionization is identified as
localization in a purely chaotic system or stabilization when the
classical phase space is mixed.

\begin{figure}[h]
  \includegraphics[width=5.5cm,angle=-90]{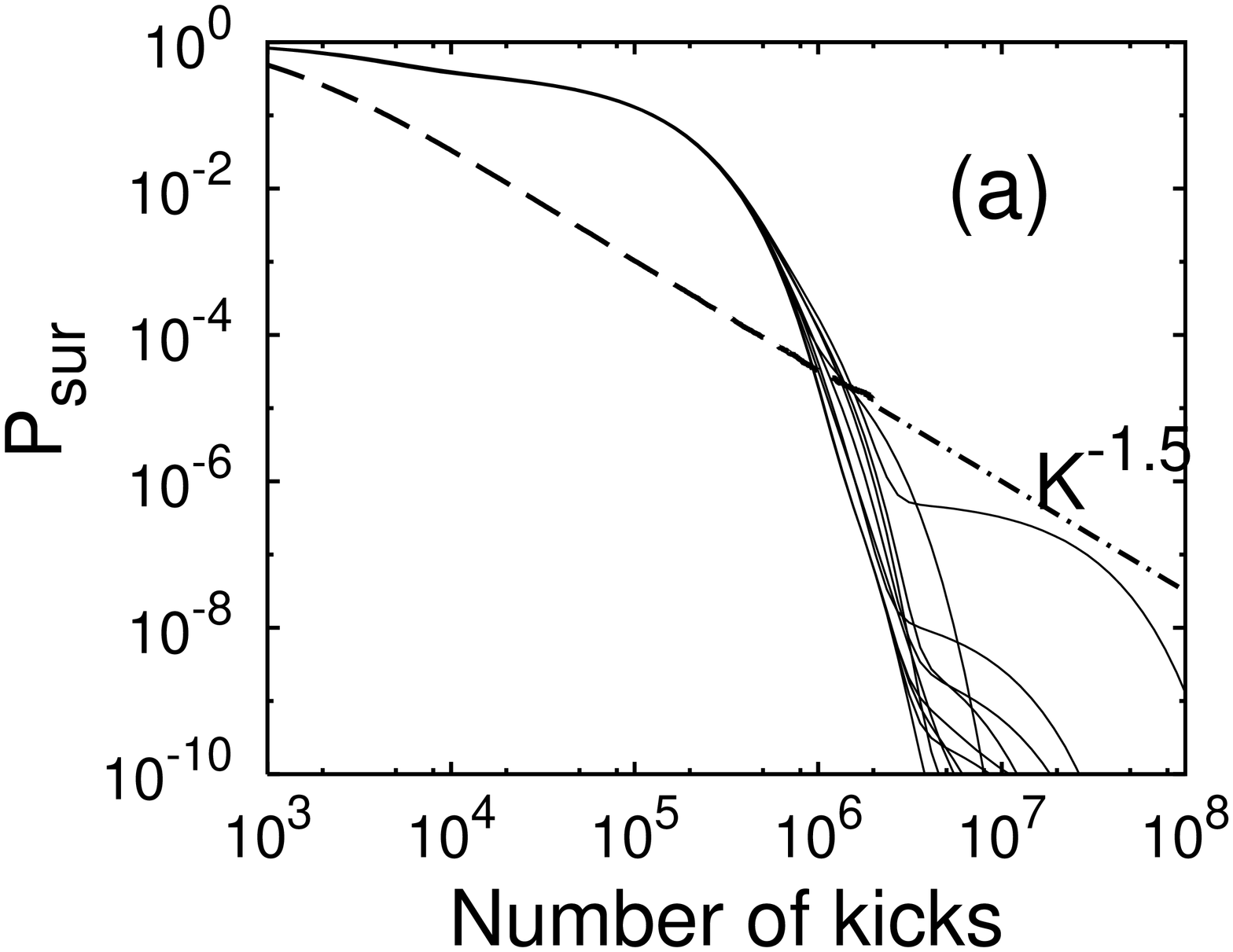}
  \includegraphics[width=5.5cm,angle=-90]{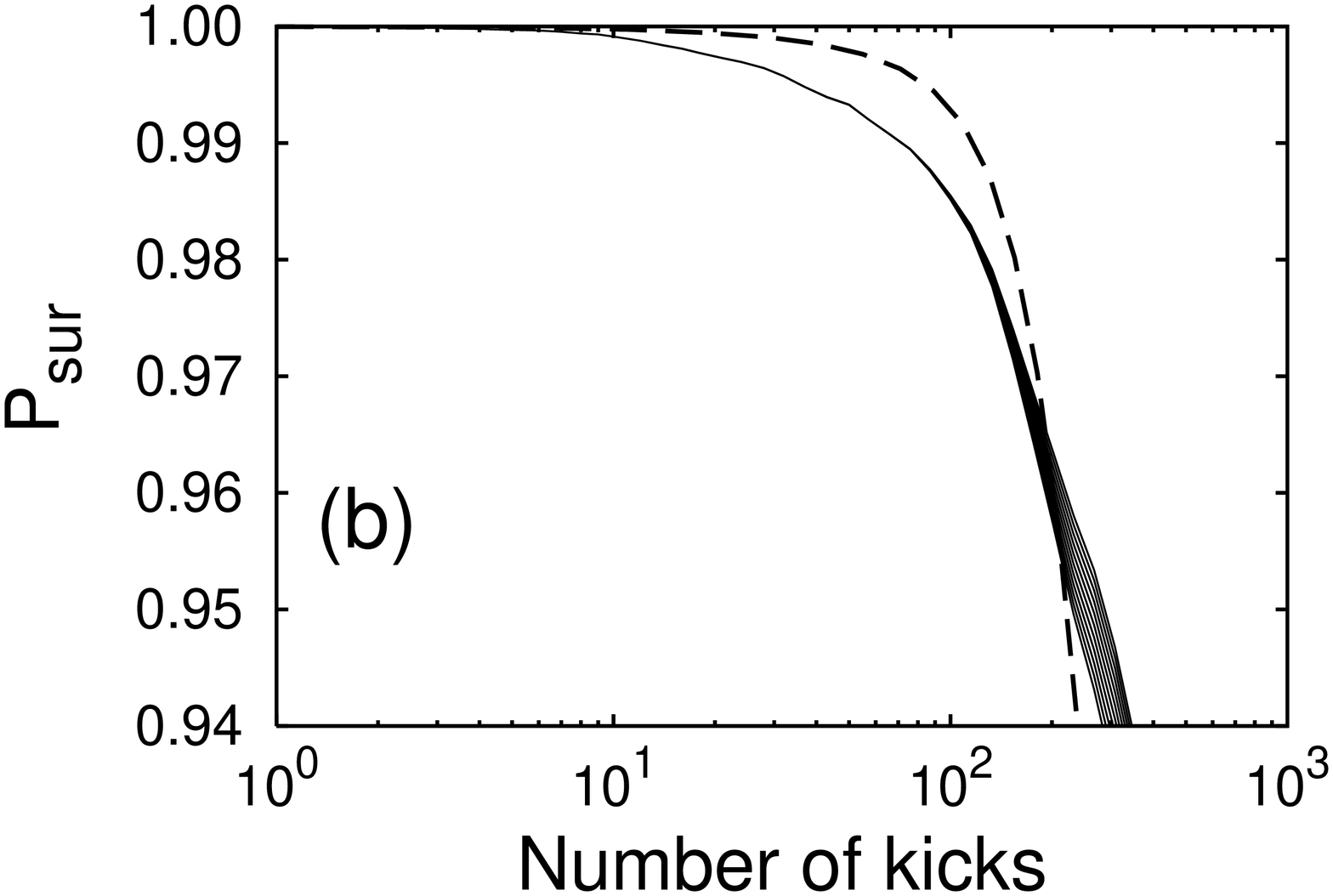}
\caption{
  Classical $P_{\rm sur}^{\rm cl}$ (dashed line) and quantum $P_{\rm
    sur}^{\rm qm}$ (solid lines) survival probability for the
  positively kicked Rydberg atom as a function of number $K$ of kicks
  for $|F_0^{\rm av}|=0.005$ ($\Delta p_0\approx 0.021$). The quantum
  data are shown for 11 frequencies $\nu_0$ uniformly distributed
  between 1.45 and 1.45008, $n_i=50$.  {\bf (a):} Long-time behavior,
  the dashed-dotted line indicates a fit to a power-law with exponent
  $\alpha=-1.5$, Eq.~(\ref{eq:algebraic}), and {\bf (b):} short-time
  behavior. }
\label{fig:psurK}
\end{figure}

\begin{figure}[h]
  \includegraphics[width=5.5cm,angle=-90]{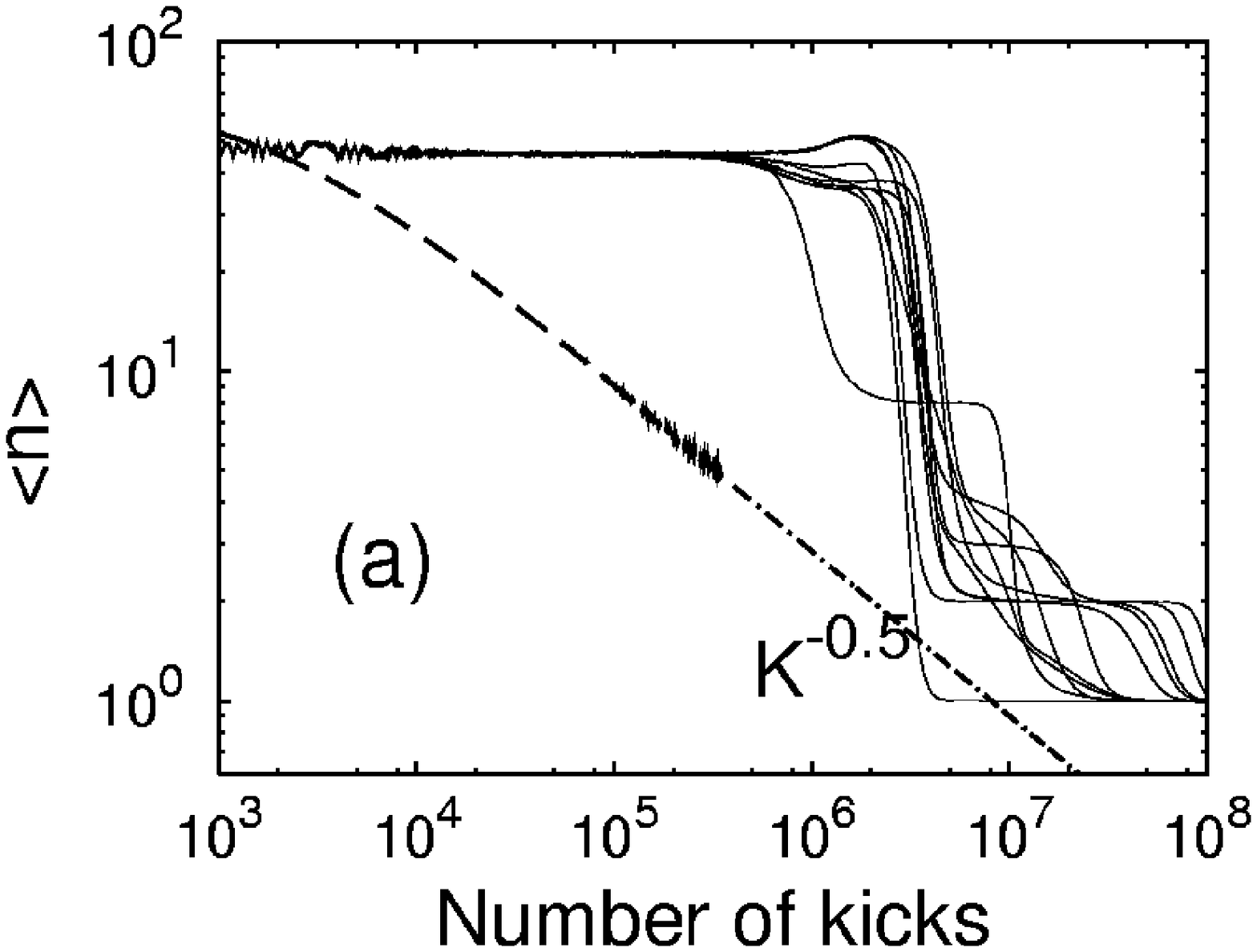}
  \includegraphics[width=5.5cm,angle=-90]{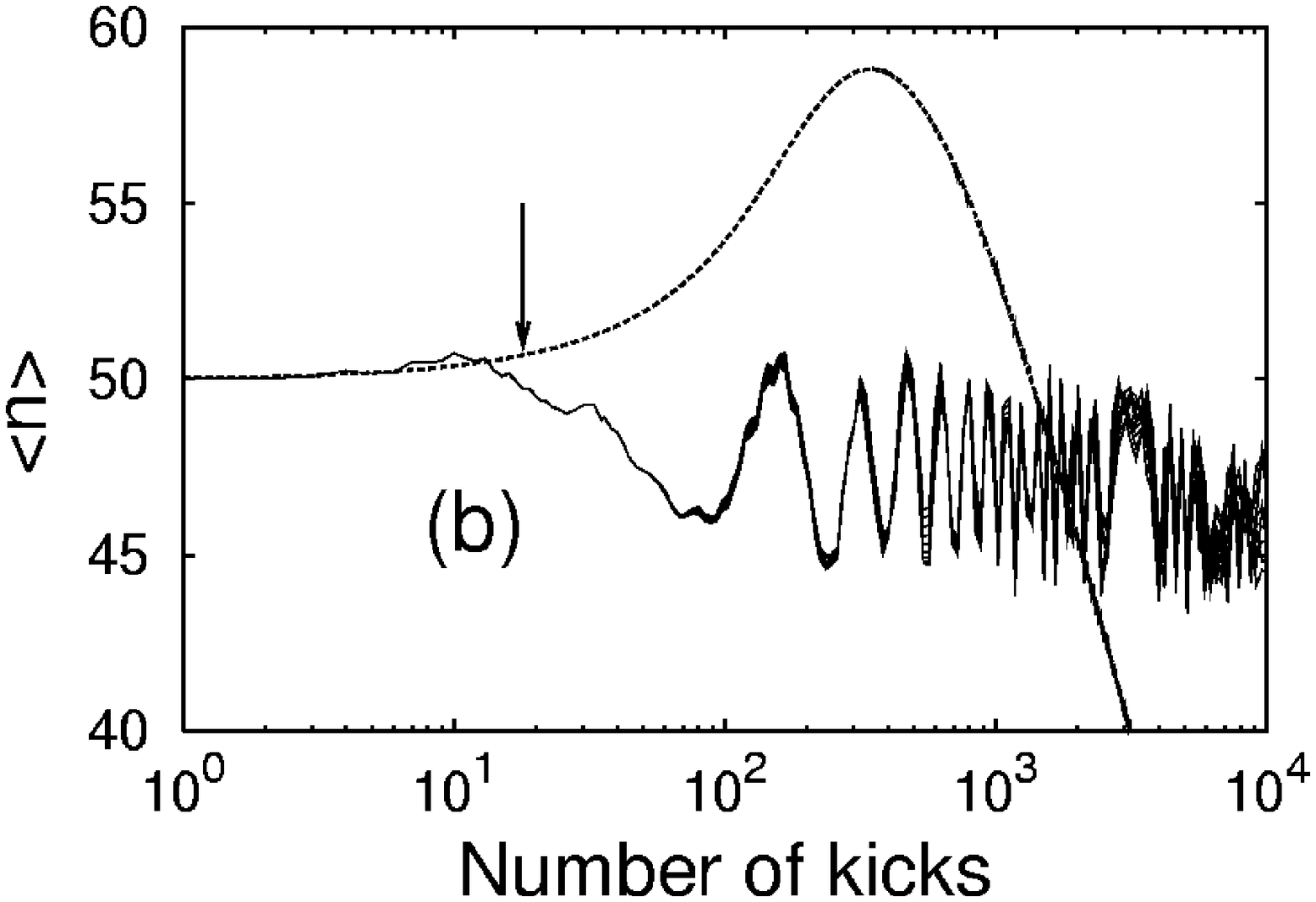}
\caption{
  Mean principal quantum number $\langle n \rangle$,
  Eq.~(\ref{eq:neff}), for the same parameters as
  Fig.~\ref{fig:psurK}, $\langle n \rangle_{\rm cl}$: dashed line;
  $\langle n \rangle_{\rm qm}$: solid lines.  {\bf (a):} Long-time
  behavior (of $\log\langle n\rangle$), the dashed-dotted line
  indicates a fit to a power-law, and {\bf (b):} short-time behavior
  of $\langle n\rangle$. The arrow in (b) indicates an approximate 
  localization time defined in Eq.~(\ref{eq:taul}). }
\label{fig:neffK}
\end{figure}

The classical $P_{\rm sur}^{\rm cl}(K)$ is obtained by the classical
trajectory Monte Carlo method (CTMC) with a micro-canonical ensemble
to represent the initial state \cite{arbi,ctmc}.  The classical phase
space (for $\Delta p>0$) is fully chaotic without any tori left
intact. We show an example for the survival probability in
Fig.~\ref{fig:psurK}. The classical $P_{\rm sur}^{\rm cl}(K)$ for
times larger than $K_0$ kicks decays algebraically \cite{hill},
\begin{equation}
P_{\rm sur}^{\rm cl}(K) = (K/K_0)^{-\alpha}
\label{eq:algebraic}
\end{equation}
with $\alpha\approx 1.5$. Both the time $K_0$ and the factor $\alpha$
are only weakly dependent on the parameters of the field ($\nu_0$ and
$\Delta p_0$). By contrast, the quantum survival probability $P_{\rm
  sur}^{\rm qm}(K)$ displays a very different and much more complex
behavior. One intriguing feature is an extreme sensitivity to the
driving frequency $\nu_0$, especially in the long-time limit (compare
also to the results in e. g.  \cite{cas87,buch98,fish}). The resulting
fluctuations in the kicked atom will be studied below in more detail
(see Sect.~\ref{sec:fluctuations}).

\begin{figure}[h]
\includegraphics[width=8cm,angle=0]{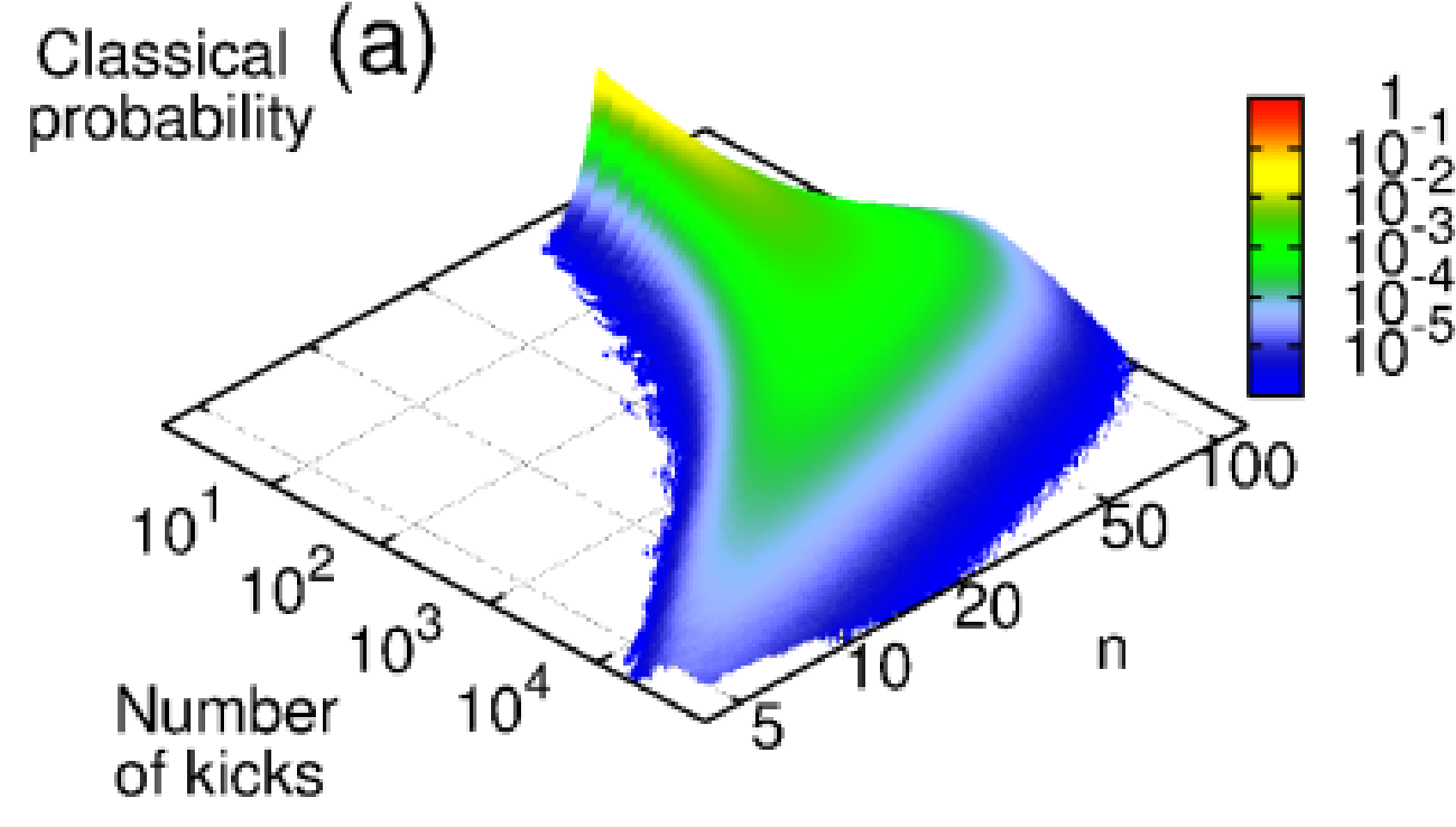}
\includegraphics[width=5.5cm,angle=-90]{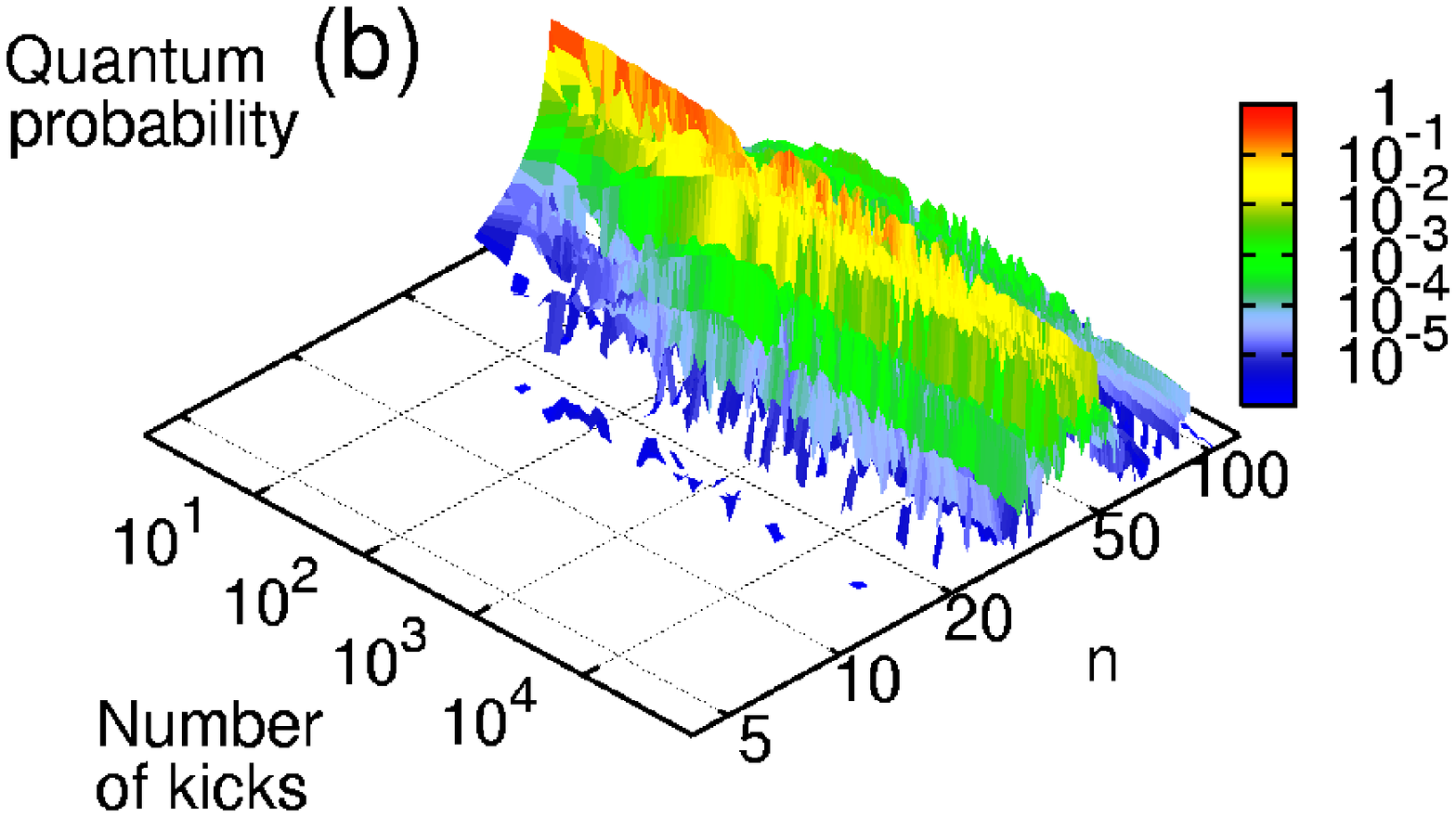}
\caption{
  (Color online) Time-dependent spectral distribution $P(n,t)$, both
  classically (a) and quantum mechanically (b) for the same parameters
  as Fig.~\ref{fig:neffK} (data shown for $\nu_0=1.45$).  }
\label{fig:ndist}
\end{figure}

For characterization of quantum (de)localization it is useful to
introduce another observable that describes the bound portion of the
wave packet.  Inspired by the relation $n=1/\sqrt{-2E}$ for bound
states, we introduce a mean quantum number as
\begin{equation}
\langle n(K) \rangle =\langle\psi(KT) | P_{\rm bound} \; 
\frac{1}{\sqrt{-2H_0}} \; P_{\rm bound}| \psi(KT)\rangle/P_{\rm sur}(K) \; .
\label{eq:neff}
\end{equation}
In this equation, the expectation value is calculated for the portion
of the wavefunction residing in the bound subspace. Equation
(\ref{eq:neff}) characterizes the specific position of that part of
the wavefunction that remains bound, in particular that remains
localized or stabilized near the initial state (or initial torus).
Indeed, $\langle n \rangle$ displays a characteristically different
behavior classically $\langle n \rangle_{\rm cl}$ and quantum
mechanically $\langle n \rangle_{\rm qm}$, both at short times
(Fig.~\ref{fig:neffK}b) and long times (Fig.~\ref{fig:neffK}a).  The
sharp drop of $\langle n \rangle_{\rm qm}$ from $50$ to $1$ at
$K\approx 10^6$ will be analyzed in Sect.~\ref{sect:longtime}.  The
corresponding non-normalized distributions $P(n,t)$ whose mean values
are given by $\langle n \rangle$ are depicted in Fig.~\ref{fig:ndist}.
The quantum $P^{\rm qm}(n,t)$ remains well localized close to the
initial state while the time evolution of $P^{\rm cl}(n,t)$ features a
rapid increase of width in the energy distribution and a shift towards
increasingly lower energies (i.e. smaller $n$).

\begin{figure}[h]
  \includegraphics[width=5.5cm,angle=-90]{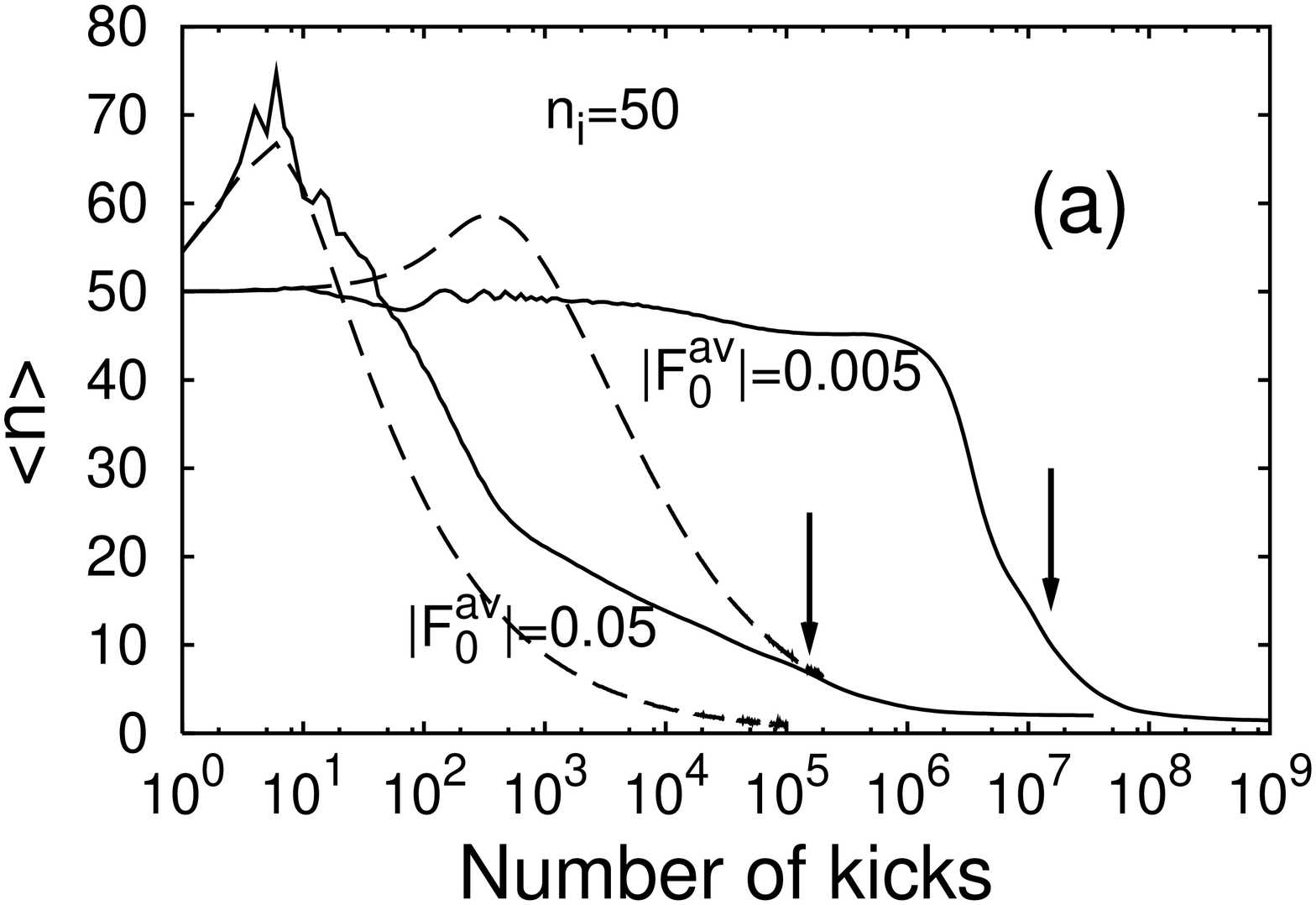}
  \includegraphics[width=5.5cm,angle=-90]{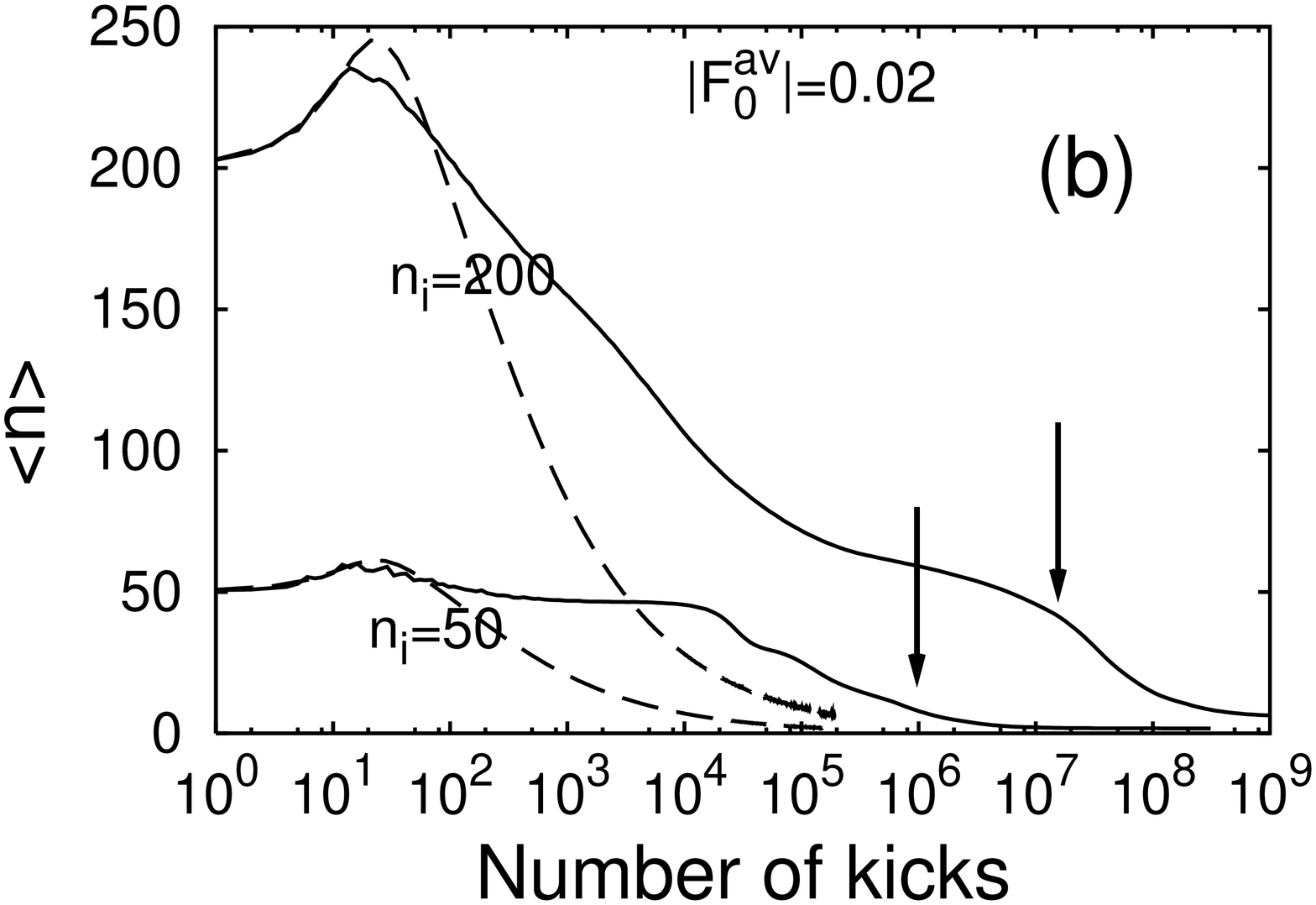}
\caption{
  $\langle n \rangle$ averaged over the interval
  $1.45\le\nu_0\le1.47$.  In {\bf (a)}, $n_i=50$ and in {\bf (b)},
  $|F_0^{\rm av}|=0.02$.  Full line: quantum result and dashed line:
  classical result.  The arrows indicate the lifetimes of the ground
  state estimated by direct coupling to the continuum. The average is
  calculated on the logarithmic scale, i.e. shown is
  $10^{<\log_{10}n>}$. }
\label{fig:neffave}
\end{figure}

\begin{figure}[h]
  \includegraphics[width=5.5cm,angle=-90]{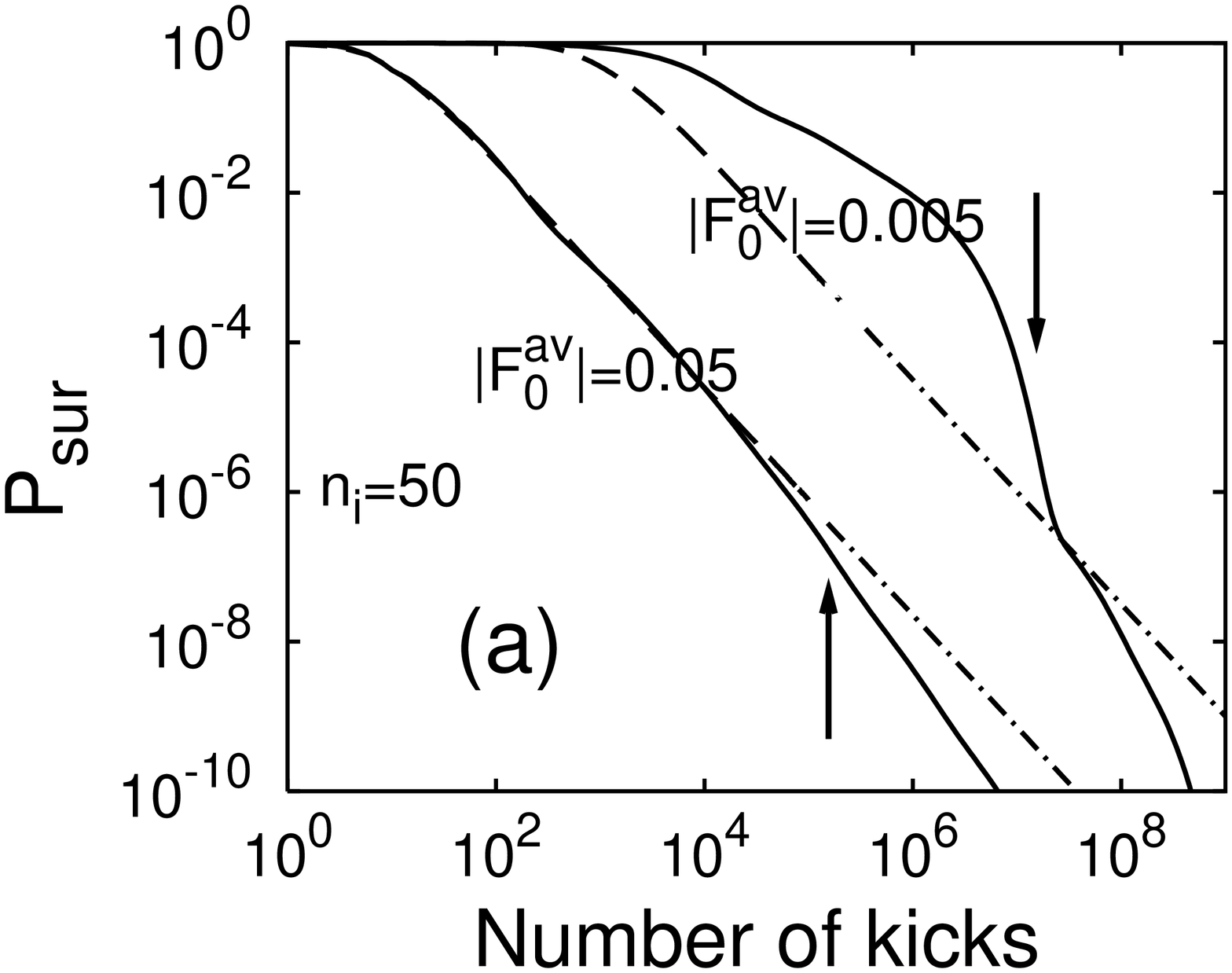}
  \includegraphics[width=5.5cm,angle=-90]{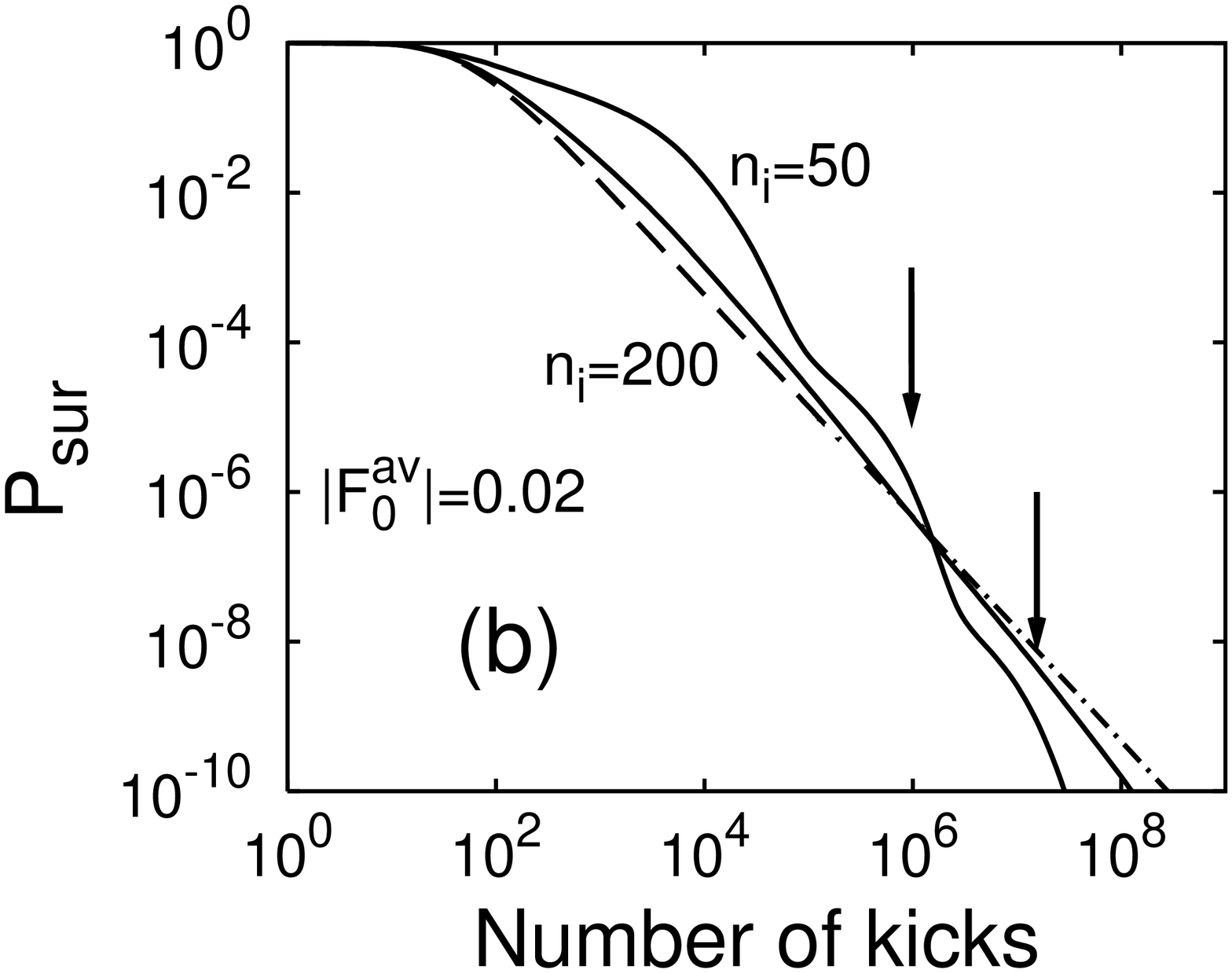}
\caption{
  $P_{\rm sur}$ averaged over the interval $1.45\le\nu_0\le1.47$.  In
  {\bf (a)}, $n_i=50$, in {\bf (b)}, $|F_0^{\rm av}|=0.02$.  Full
  line: quantum result, dashed line: classical result and
  dashed-dotted line: fit of classical result to a power-law.  The
  arrows indicate the lifetimes of the ground state estimated by
  direct coupling to the continuum.  The average is calculated on the
  logarithmic scale.  }
\label{fig:tprobave}
\end{figure}

Note, that classically $n$ does not possess a lower bound $n=1$, as
the quantum distribution does. The fact that the energy distribution
can grow to negative values without bound is at the root cause for the
algebraic decay.  The probability for crossing the $E=0$ line (i.e.
the $n=\infty$ line) in the next time step and thus for reducing
$P_{\rm sur}^{\rm cl}$ decreases as the mean distance to the $E=0$
line as measured by $\langle n(K) \rangle_{\rm cl}$ increases with
$t$.  We now relate the power-law decay Eq.~(\ref{eq:algebraic}) to
the motion of $P^{\rm cl}(n,t)$ away from the threshold.  The
classical survival probability for large times is well fitted by
Eq.~(\ref{eq:algebraic}) with $K_0\approx 1000$. The scaled average
quantum-number is also well described by a power-law,
\begin{equation} 
\langle n_0 (K)~\rangle_{\rm cl} = b(K/K_0)^{-0.5} \;, 
\label{eq:neffpowerlaw}
\end{equation}
where the power-law decay is assumed to start at the same time as that
of the survival probability and $b\approx 1.8$.  The differential form
of Eq.~(\ref{eq:algebraic}),
\begin{equation}
\frac{d P_{\rm sur}^{\rm cl}(K)}{dK}=
-\frac{\alpha}{K} P_{\rm sur}^{\rm cl}(K) \; ,
\label{eq:algebraicdiff}
\end{equation}
can now be expressed by means of Eq.~(\ref{eq:neffpowerlaw}) as
follows,
\begin{equation}
\frac{d P_{\rm sur}^{\rm cl}(K)}{dt}
=-\frac{d}{K_0}\langle n_0(K)~ \rangle_{\rm cl}^2
P_{\rm sur}^{\rm cl} 
\label{eq:powerdiff}
\end{equation}
with $d=\alpha/b^2$.  For some cases with stronger average fields
($F_0^{\rm av}=0.02$ and $0.05$) shown in Figs.~\ref{fig:neffave} and
\ref{fig:tprobave} (where $K_0\approx 60$ and $9$, respectively) we
find that the power-law for $\langle n_0 \rangle_{\rm cl}$,
Eq.~(\ref{eq:neffpowerlaw}), holds with approximately the same
constant $b$.  Hence, Eq.~(\ref{eq:powerdiff}) is still applicable and
the constant $d$ has essentially the same value.

We now compare the numerical signatures of quantum localization for
intermediary times as seen in $P_{\rm sur}$, Figs.~\ref{fig:psurK}a
and \ref{fig:tprobave}, to the signatures seen in $\langle n \rangle$,
Figs.~\ref{fig:neffK}a and \ref{fig:neffave}.  In all cases studied,
the $\langle n \rangle^{\rm qm}>\langle n \rangle^{\rm cl}$ and a
clear discrepancy between the quantum and the classical value can be
seen for weak fields and low $n_i$.  The discrepancy decreases as
$|F_0^{\rm av}|$ or $n_i$ increases, i.e. $\langle n \rangle^{\rm qm}$
approaches the classical limit, however quite slowly. The approach to
the classical limit appears faster in $P_{\rm sur}$. Here signatures
of quantum localization can only be see for weak fields and low $n_i$
while in the other cases shown, $P_{\rm sur}^{\rm qm}$ has already
reached the classical limit.  We thus conclude that the more locally
focused $\langle n \rangle$ provides stronger signature of quantum
localization, i.e. the suppression of diffusion away from the initial
state, than the $P_{\rm sur}$ probing the entire bound part of phase
space.

\subsection{Short-time dynamics: the cross-over to quantum localization}
\label{sec:shortt}

We take now a closer look at the short time dynamics for the weak
field strength, Fig.~\ref{fig:psurK}b and Fig.~\ref{fig:neffK}b. Here
a first surprise appears.  One would, generally, expect $P_{\rm
  sur}^{\rm cl}$ and $P_{\rm sur}^{\rm qm}$ to agree with each other
up to the localization time (or quantum break time, also referred to
as Ehrenfest time) $\tau_l$. The present case is non-generic in that
the classical phase space distribution remains up to $\tau_l$
($\approx 200$ kicks) more localized when one identifies $P_{\rm sur}$
as a measure for localization.  This quantum enhancement of ionization
takes place even though the classical $\langle n \rangle_{\rm cl}$
moves closer to the ionization threshold while the quantum $\langle n
\rangle_{\rm qm}$ remains close to the initial vale
(Fig.~\ref{fig:neffK}b).  The origin is a true quantum effect:
perturbative single photon absorption of high frequency $m\nu_0$ from
higher-harmonic components with $m\ge m_c(n_i)$ sufficient to reach
the continuum. Here
\begin{equation}
m_c(n_i)=n_i/(2\nu_0)
\label{eq:mc} \; .
\end{equation}

A border for the field strength at which a single kick driving a
high-lying Rydberg state displays quantum-classical correspondence can
be found as follows \cite{jbhcp}: The average classical energy
transfer from a kick is
\begin{equation}
\Delta E_k = \Delta p_0^2  \; .
\end{equation}
This energy transfer can be resolved quantum mechanically if $\Delta
E_k$ is not smaller than the quantum energy spacing $\Delta
E_{n_i}^0=1/{n_i}$, leading to the critical momentum transfer
\begin{equation}
\Delta p_0^{\rm crit}=1/\sqrt{n_i}.
\label{eq:dpcrit}
\end{equation}
For $\Delta p_0 > \Delta p_0^{\rm crit}$, the quantum and classical
distributions after a single kick agree all the way up to the
threshold $E=0$ since the quantum level spacing decreases as
$n\to\infty$, and a good agreement between the quantum and classical
survival probabilities is achieved. The difference between the
classical and quantum distributions for $\Delta p_0 < \Delta p_0^{\rm
  crit}$ can be understood by considering the dipole limit $\Delta p_0
< \Delta p_0^{\rm dipole} = 1/n_i$, for which the transition operator
for a single kick, $\exp(i\Delta p\;q)$, reduces to
\begin{equation}
\exp\left(i\Delta p\;q\right)\approx 1+i\Delta p_0 q_0 n_i+ ... \;.
\label{eq:dipole}
\end{equation}
In this limit, the quantum transition amplitude is proportional to
$\Delta p$ and the probability is proportional to $\Delta p^2$. 
By contrast, the classical probability is proportional to 
$\Delta p^5$ implying that the quantum survival probability after 
one kick is smaller than the classical one \cite{jbhcp}.
Physically, this can be understood by 
considering the Fourier transform of a delta kick,
\begin{eqnarray}
\delta(t-t_0)=\frac{1}{2\pi}\int\limits_\infty^\infty d\omega
\exp\left(i\omega(t-t_0)\right) \; ,
\end{eqnarray}
indicating a ``white'' spectrum.
The quantum system can absorb (virtual) photons of arbitrarily
high frequency from the white spectrum accompanied by only a small
momentum transfer $\Delta p_0\ll 1$. Processes with large energy 
transfer $\Delta\omega_0\gg 1$ but small momentum transfer $\Delta p_0$
are far from the
line  $\Delta \omega_0 \approx \Delta p_0^2/2$ in the 
$\Delta\omega_0\;-\;\Delta p_0$
dispersion plane for a quasi-free electron,
and thus effectively is inaccessible for a classical momentum transfer 
process. Only for deeply bound electrons with large local orbital 
momentum, $p_0$, near the nucleus, with  
$\Delta\omega_0\approx p_0 \Delta p_0 \gg \Delta p_0^2/2$ can such 
processes occur in the classical case. 
The density of classical phase space points
for the initial state with such high  $p_0$ is, however, very small.
In the quantum case, the high-frequency components in the driving 
field interact 
non-locally with the whole initial state
leading to an enhanced ionization probability. 
This corresponds to dipole allowed transitions.
Classical-quantum correspondence
is only restored when classical diffusion in phase space dominates
the quantum enhancement due to (virtual) photon absorption.

\begin{figure}[t]
  \includegraphics[width=5.5cm,angle=-90]{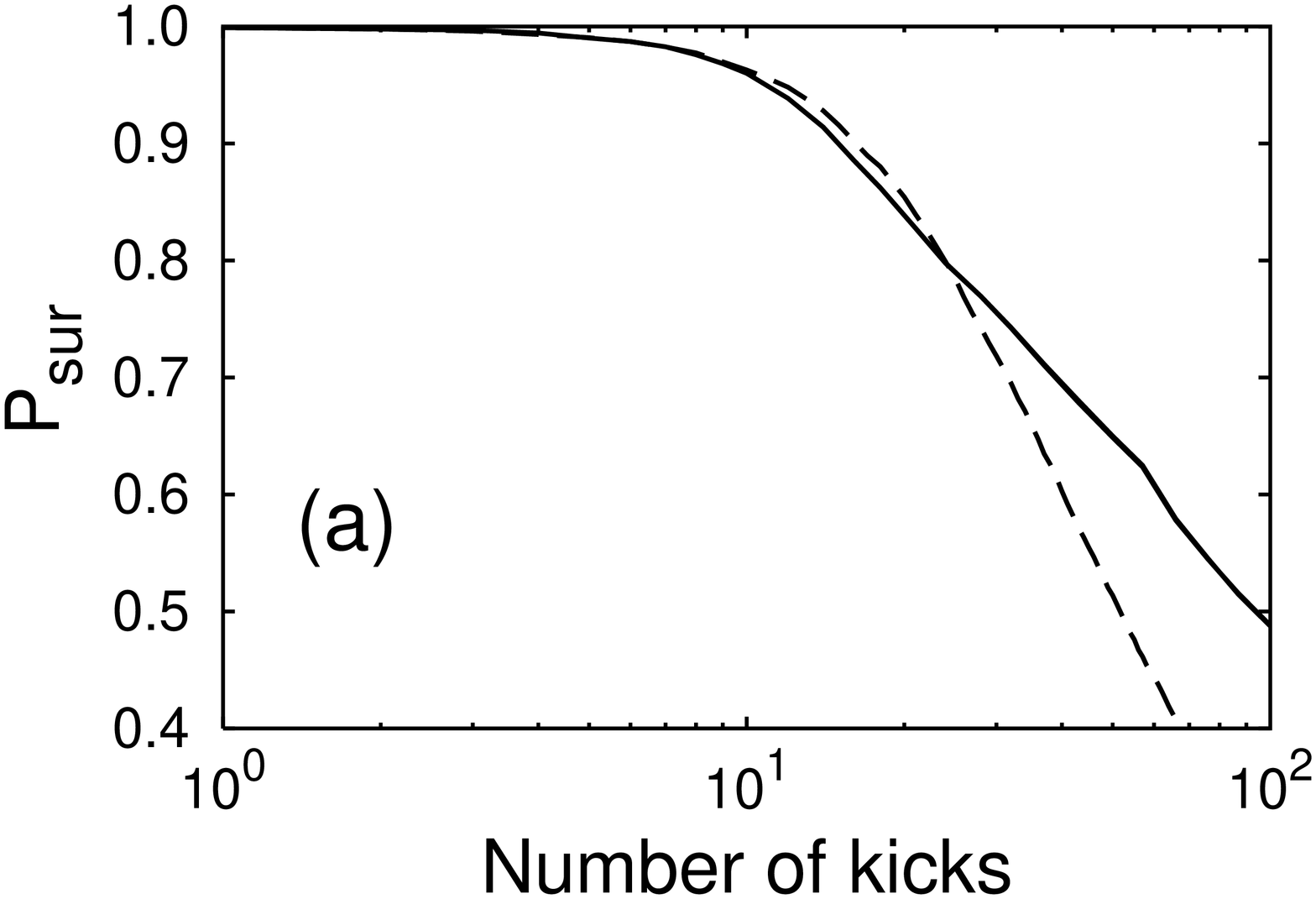}
  \includegraphics[width=5.5cm,angle=-90]{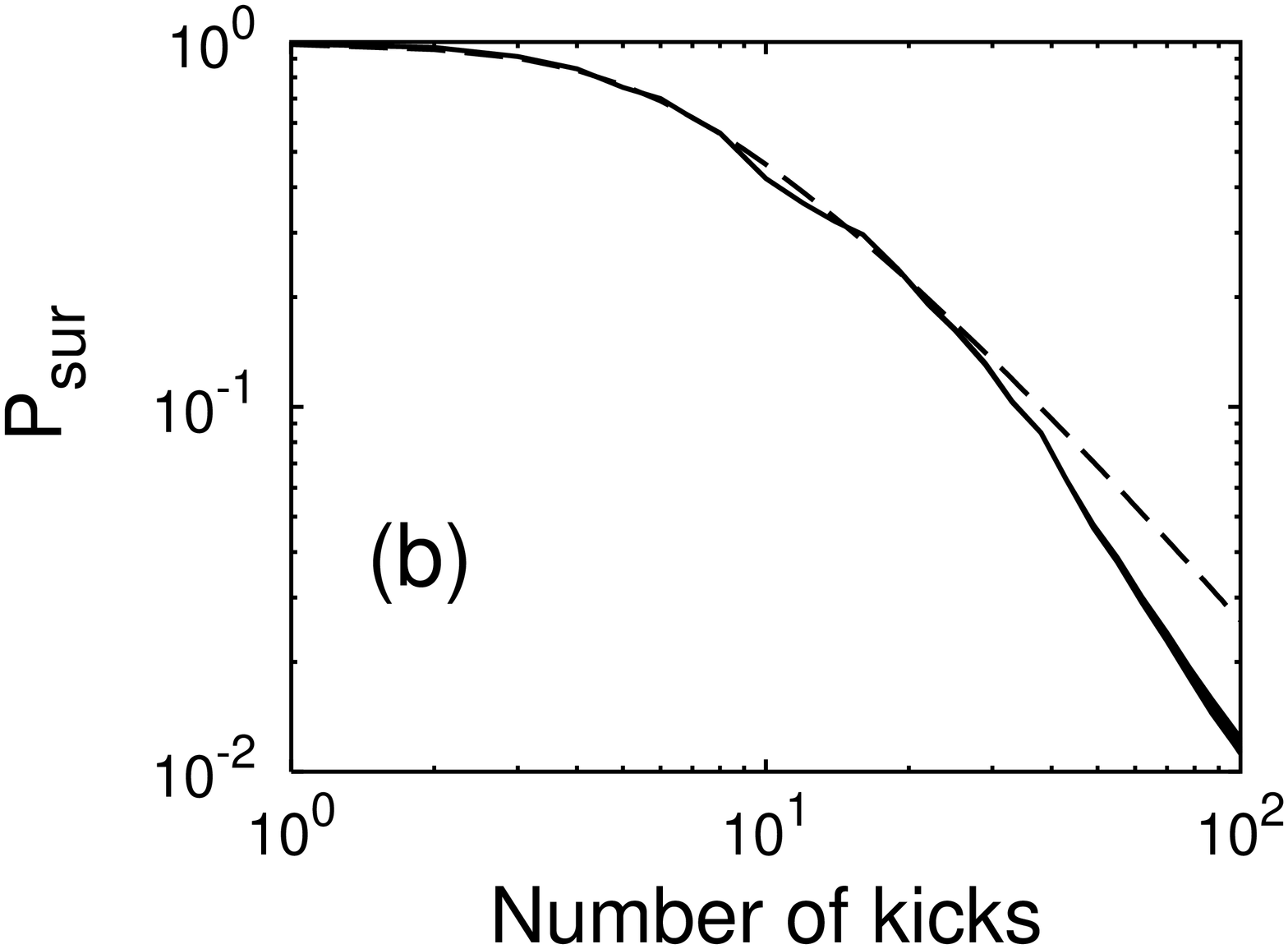}
  \includegraphics[width=5.5cm,angle=-90]{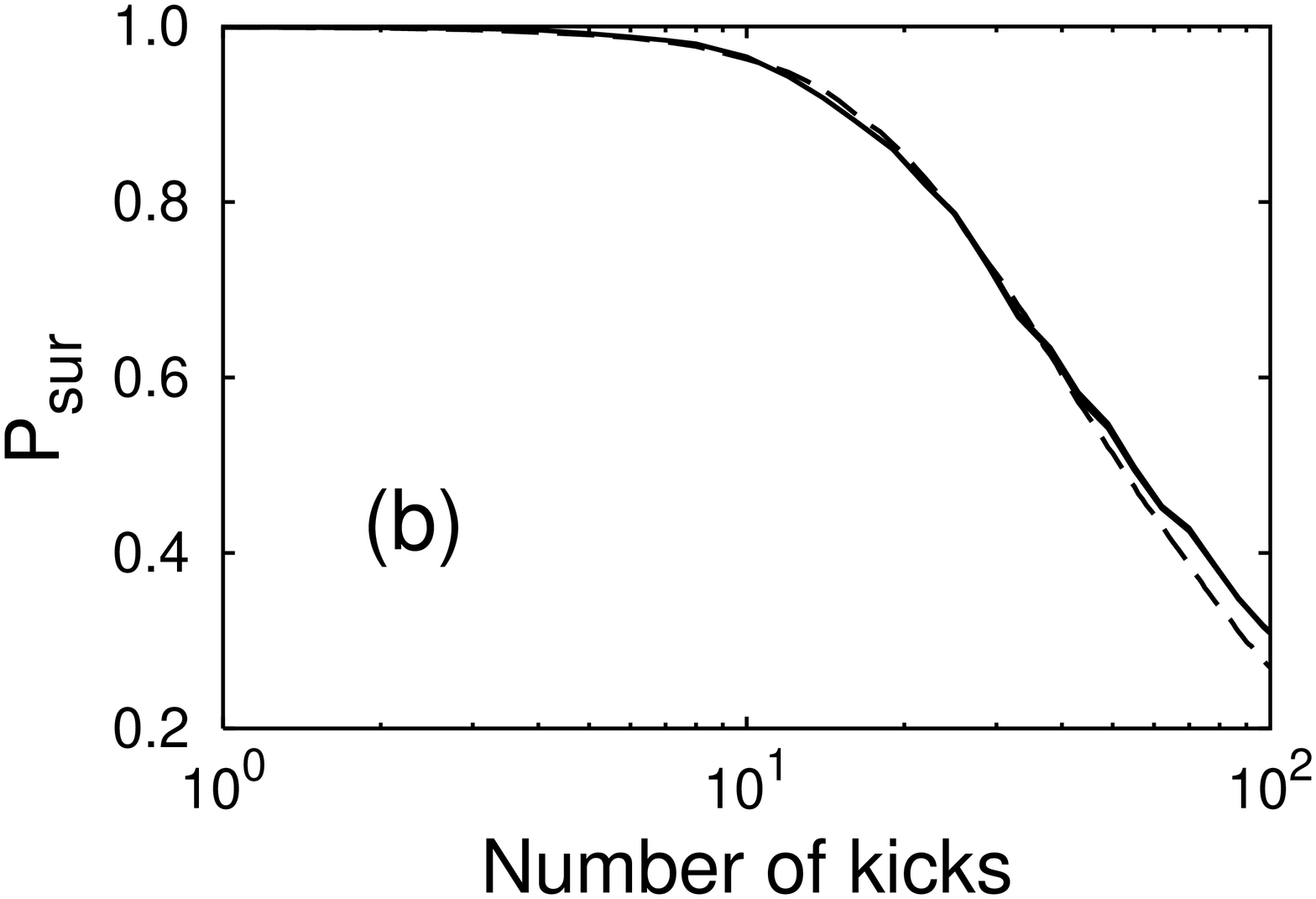}
\caption{
  As Fig.~\ref{fig:psurK}b, however for {\bf (a):} $|F_0^{\rm
    av}|=0.02$ and $n_i=50$, {\bf (b):} $|F_0^{\rm av}|=0.05$ and
  $n_i=50$, and {\bf (c):} $|F_0^{\rm av}|=0.02$ and $n_i=200$.  }
\label{fig:psurdiff}
\end{figure}

For $n_i=50$ and $\nu_0=1.45$ the corresponding critical fields are
$|F_0^{\rm crit}|=0.033$ and $|F_0^{\rm dipole}|=0.0047$, i.e. the
case studied in Figs.~\ref{fig:psurK} and \ref{fig:neffK} are close to
the dipole limit. The important point is that the dynamical role of
the high-harmonics spectrum of (virtual) photons is responsible for
the surprising, non-generic features of the periodically kicked
Rydberg atom, different from other systems such as the Rydberg atom in
a microwave field. This is directly verified by increasing the field
such that the scaled momentum transfer $\Delta p_0\approx \Delta
p_0^{\rm crit}$, see Fig.~\ref{fig:psurdiff}.  We show first a case in
the transition region ($|F_0^{\rm av}| =0.02$, $n_i=50$) where still
some traces of the non-generic features at short times can still be
seen.  For larger $n_i$ or larger $|F_0^{\rm av}|$ such that $\Delta
p_0 > \Delta p_0^{\rm crit}$ the quantum-classical agreement for short
times is quite well fulfilled while discordance is found only for
larger times, $t>\tau_l$.  The short-time dynamics now conforms with
the naive expectation of close classical-quantum correspondence for
all $t< \tau_l$.

\begin{figure}[h]
  \includegraphics[width=5.5cm,angle=-90]{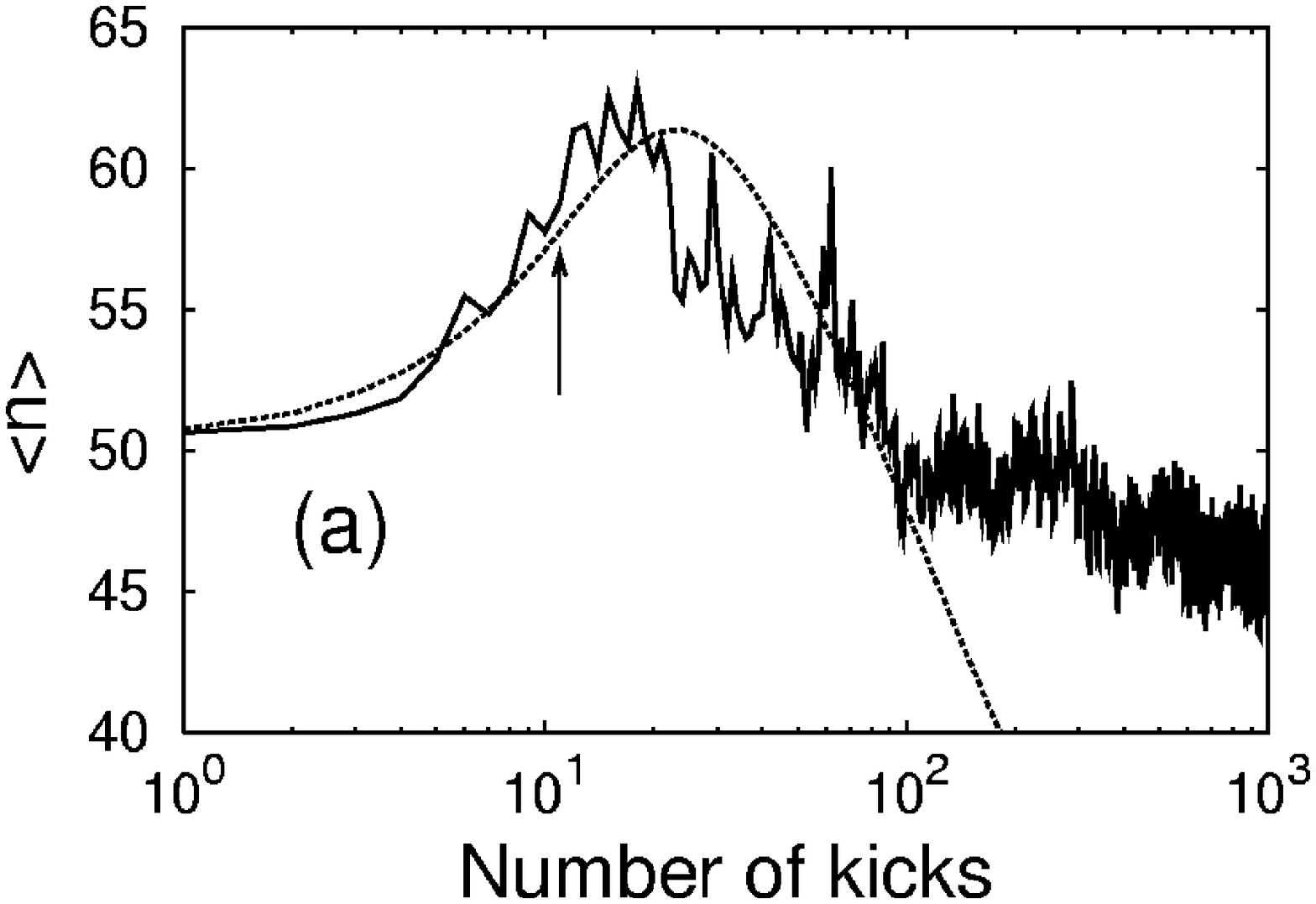}
  \includegraphics[width=5.5cm,angle=-90]{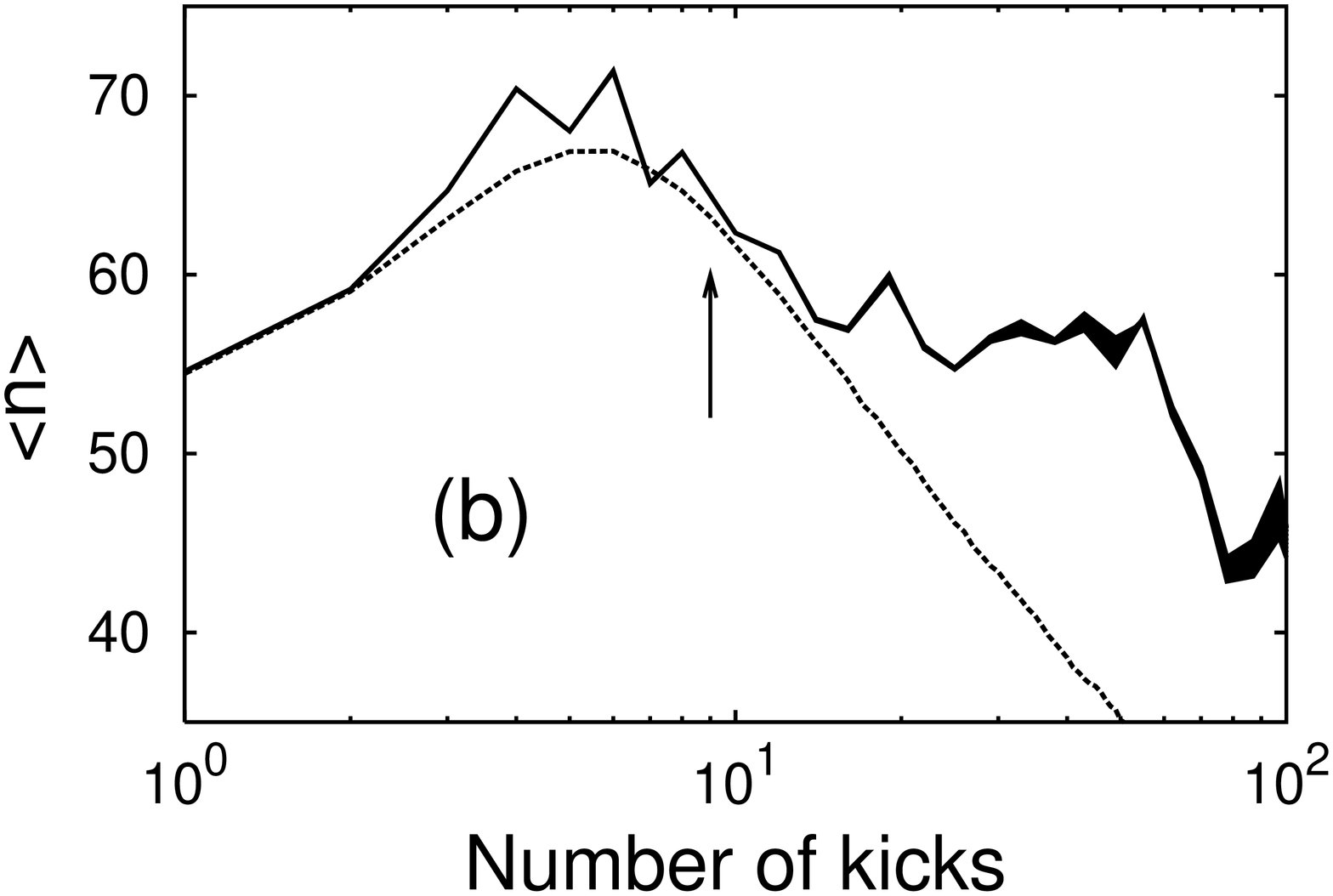}
  \includegraphics[width=5.5cm,angle=-90]{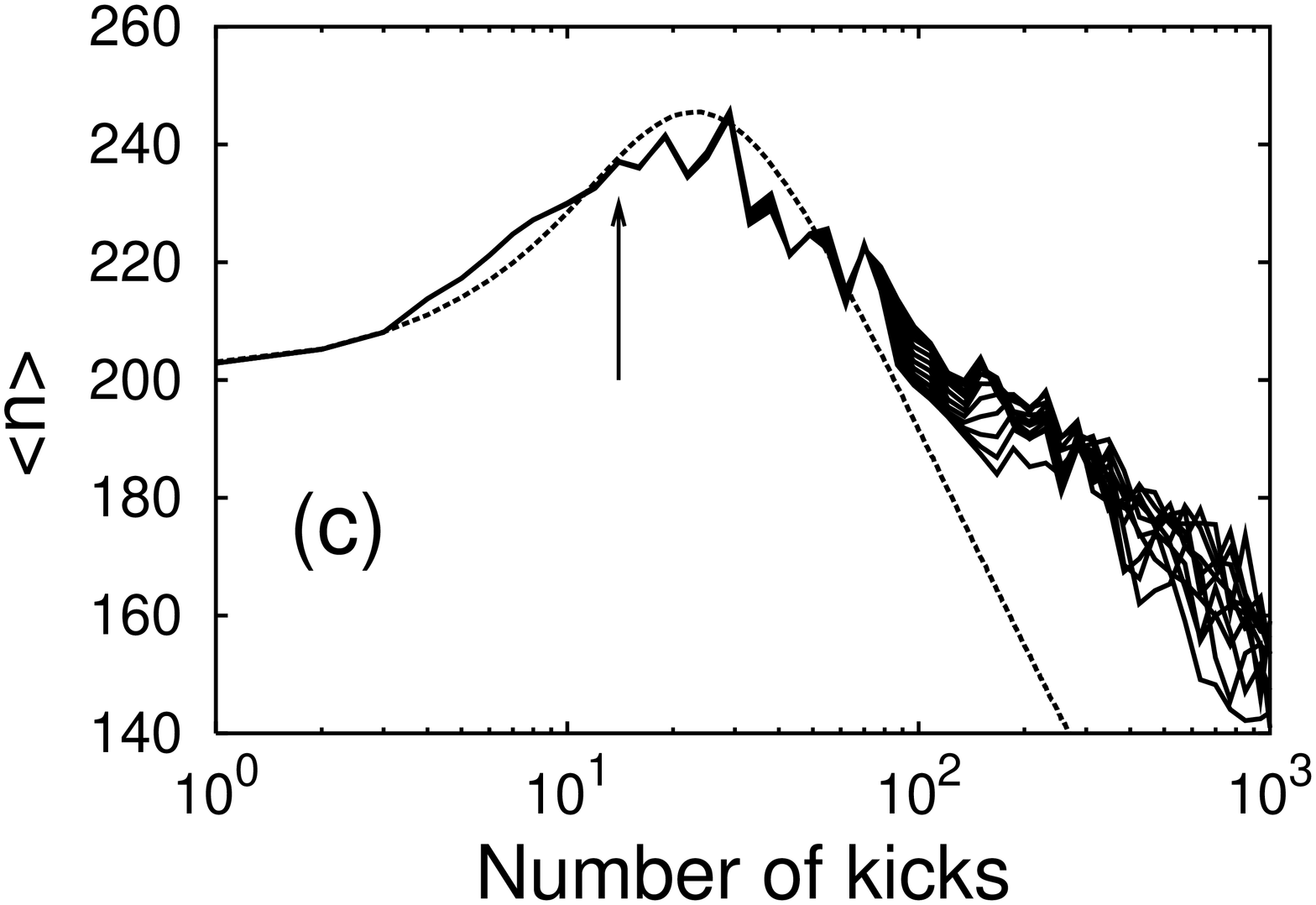}
\caption{
  As Fig.~\ref{fig:neffK}b, however for {\bf (a):} $|F_0^{\rm
    av}|=0.02$ and $n_i=50$, {\bf (b):} $|F_0^{\rm av}|=0.05$ and
  $n_i=50$, and {\bf (c):} $|F_0^{\rm av}|=0.02$ and $n_i=200$.
  The arrows indicate an approximate localization time defined in 
Eq.~(\ref{eq:taul}).  }
\label{fig:neffdiff}
\end{figure}

Well-known arguments \cite{aleiner} for the localization time lead to
the order of magnitude estimate
\begin{equation}
\tau_l\approx \frac{1}{\langle L \rangle}\; 
\ln\left(\frac{\langle q \rangle }{\lambda_i}\right) \; \; ,
\label{eq:taul}
\end{equation}
where $\langle q \rangle=n_i^2$ is the typical spatial extension of
the initial state, $\lambda_i=n_i$ the wave length of the initial
state, and $\langle L \rangle$ the mean Lyaponov exponent.  
$\tau_l$ is indicated by arrows in Fig.~\ref{fig:neffK}b
and Fig.~\ref{fig:neffdiff}. For $\langle n \rangle$,
$\tau_l$ gives a reasonable estimate for the number of kicks up to where
quantum and classical dynamics mirror each other.

The situation for the survival probability is more involved: For
$\Delta p_0 < \Delta p_0^{\rm crit}$, Eq.~(\ref{eq:dpcrit}) (see
Figs.~\ref{fig:psurK} and \ref{fig:psurdiff}a), no agreement between
$P_{\rm sur}^{\rm cl}$ and $P_{\rm sur}^{\rm qm}$ is found for short
times and a break time $\tau_l$ now defined as the time where $P_{\rm
  sur}^{\rm cl}$ gets smaller than $P_{\rm sur}^{\rm qm}$ due to
quantum localization takes on a different meaning. Consequently, this
break time is not well described by Eq.~(\ref{eq:taul}).  For the
cases with $\Delta p_0 > \Delta p_0^{\rm crit}$,
(Figs.~\ref{fig:tprobave} and \ref{fig:psurdiff}b and c) we find no
clear numerical signatures of localization in $P_{\rm sur}$.

\subsection{Long-time dynamics: the second cross-over}
\label{sect:longtime}

Quantum localization in the kicked Rydberg atom is transient.  After a
large but finite number of kicks (Figs.~\ref{fig:psurK}b and
\ref{fig:psurdiff}), the quantum survival probability rapidly
decreases and at a second cross-over time (delocalization time)
$\tau_D$ falls below even the classical value.  Simultaneously,
$\langle n\rangle_{\rm qm}$ drops sharply from values near $n_i$ to 1
(Figs.~\ref{fig:neffK}b and \ref{fig:neffdiff}b).  Beyond the
cross-over point, $\langle n\rangle_{\rm cl}$ falls to values well
below unity inaccessible to quantum mechanics.  There, the residual
fraction of the classical phase is ``sheltered'' and continues to
decay slowly, i.e.  algebraically.  By contrast, the quantum
bound-state probability decays exponentially in the long-time limit.
Beyond $\tau_D$ the slow classical algebraic decay ``wins'' over the
exponential decay.  This novel scenario is markedly different from the
sinusoidal driven Rydberg atom, where the quantum transport is bounded
from below because of the regular phase space region, or, in a quantum
picture, because of vanishingly small transition probabilities 
to states with low principal quantum numbers \cite{cas87,koch,saw02}.

The rate of the long-time decay and, thus, of $\tau_D$ can be
estimated from the decay rate of the state $n=1$ due to the high
harmonics.  This is the lowest-lying and most stable component of any
coherent superposition forming a Floquet state.  All frequencies of
the driving field Eq.~(\ref{eq:hamfourier}) with scaled energy
$m\nu_0$ large enough to couple the Stark state having the largest
overlap with the $n=1$ hydrogenic state to the Stark continuum
contribute to the transition probability according to Fermi's golden
rule,
\begin{equation}
\Gamma_m=2\pi\rho(E_m) |F^{\rm av} z(n=1,E_m)|^2 \; .
\label{eq:golden}
\end{equation}
with $E_m=-1/2+m\nu_0/n_i^3$. Both dipole matrix elements $z(n,E)$ and
the density of continuum states $\rho(E)$ are numerically obtained by
diagonalizing the Stark Hamiltonian (\ref{eq:hamstark}) in the
pseudo-spectral basis.  Summing over all contributions leads to a
delocalization time $\tau_D=\left(\sum_m\Gamma_m\right)^{-1}$ which
corresponds to within a factor of two to the lifetime of the most
long-lived Floquet state with $\langle n \rangle\approx 1$.  $\langle
n \rangle_{\rm qm}$ averaged over an interval in $\nu$ ($1.45\le \nu_0
\le 1.47$) together with the estimated $\tau_D$ are shown in
Fig.~\ref{fig:neffave}. $\tau_D$ clearly gives a good estimate of the
time-scale on which the break-down of quantum localization as seen in
$\langle n \rangle$ takes place.  The estimated $\tau_D$ also coincide
with $P_{\rm sur}^{\rm qm}$ getting smaller than the classical value
(Fig.~\ref{fig:tprobave}).  We thus attribute the break-down of
quantum-localization to the higher harmonics in the driving field.
The breakdown of localization is expected if the
ground state $n=1$ is directly coupled to the continuum.
We note parenthetically that an experimental realization would require 
half-cycle pulses with high-frequency components in the UV region.
Work is currently under way on a protocol
to produce half-cycle pulses in the atto-second regime
\cite{atto}.

We now comment on the decay of the quantum survival probability for
$|F_0^{\rm av}|=0.005$ and $0.02$, $n_i=50$, seen for intermediate
times ($K <10^6$ and $10^4$ in Figs.~\ref{fig:tprobave}a and b,
respectively).  Here $\langle n \rangle^{\rm qm}\approx n_i$
(Fig.~\ref{fig:neffave}) indicating the localization of the quantum
distribution $P(n,t)^{\rm qm}$ (see also Fig.~\ref{fig:ndist}b). As
for $\langle n \rangle$, Eq.~(\ref{eq:neff}), we attribute the decay
of the localized bound-part of the quantum wavefunction in the kicked
atom to the higher harmonics present in the driving field, directly
coupling the wave packet localized close to the initial state with the
continuum.

\section{Fluctuations in the survival probability}
\label{sec:fluctuations}

The quantum survival probability shown in Fig.~\ref{fig:psurK} display
strong fluctuations under small variations of the kick frequency. Such
fluctuations are a direct consequence of the photonic localization
scenario \cite{cas87,buch98,jensen}, to be described in the following.

\subsection{High harmonics in the localization regime}

The observation of localization, i.e. the suppression of ionization,
or, equivalently, the freezing out of portions of the wavefunction
near $n_i$, raises the question as to the underlying mechanism in the
presence of the harmonic spectrum (Eq.~(\ref{eq:hamfourier})~). High
harmonics cannot only directly couple to the continuum, see
Sect.~\ref{sect:longtime}, but they also can lead to sequential
excitation through a ladder of intermediate (quasi)bound states.  This
channel is the dominant mechanism for the excitation and ionization by
the harmonic driving by the lowest harmonic $\nu_0$.  Jensen et al.
\cite{jensen} have discussed the suppression of the sequential ladder
excitation by an harmonic driving as a mechanism for localization
(``photonic localization''). It is therefore instructive to extend
this approach to the present multi-photon case.

Following \cite{jensen} we assume quasi-resonant one-photon
transitions to dominate the time evolution, i.e. we consider only
transition between (bound) states with energy differences to the
initial state approximately equal to $k$ times the fundamental photon
energy $E_0^\gamma=\nu_0/n_i$.  The detunings
\begin{equation}
\Delta_k=E_k-E_{n_i}-k\nu 
\label{eq:detuning}
\end{equation}
for the quasi-resonant states $\mid k \rangle$ form a pseudo-random
sequence of numbers.  Using the rotating wave approximation and
setting $c_k(t)=b_k(t)\exp(-i\Delta_k t)$, with $b_k(t)$ the expansion
coefficient for the ${\rm k^{th}}$ quasi-resonant state in the
interaction picture, leads to a set of coupled differential equations,
\begin{equation}
i\frac{dc_k}{dt}=\sum\limits_{k'} H^J_{kk'} c_{k'}
\label{eq:andersondiff}
\end{equation}
with the matrix $H^J$ given by
\begin{equation}
H^J_{kk'}=\Delta_k \, \delta_{kk'}+V_{k,k'} .
\label{eq:andersonmatrix}
\end{equation}
The semiclassical expression for the coupling matrix elements is
\cite{jensen}
\begin{equation}
V_{kk'}\approx 0.411F/\left((n_k\,n_{k'})^{3/2} (m\nu_0)^{5/3}\right)
\end{equation}
with $m=|k-k'|$ and $n_k$ the main quantum number of the ${\rm
  k^{th}}$ quasi-resonant state.  While for the harmonically driven
case $V_{kk'}\ne 0$ only for $m=1$, the coupling matrix elements for
the periodically kicked Rydberg atom in the ${\rm m^{th}}$ off
diagonal are proportional to $m^{-5/3}$. Due to the randomness of the
detunings $\Delta_k$, the matrix $H^J$ is thus a pseudo-random,
power-law banded matrix. In random power-law banded matrices with the
elements decreasing as $m^{-\beta}$, the eigenstates are (weakly)
localized for $\beta\ge 1$ \cite{fyod}. Our extension of the model
introduced in \cite{jensen} on the basis of \cite{cas87} thus predicts
that the eigenstates for the kicked Rydberg atom should be localized.
Infering from the numerical observation of quantum localization that
the detunings are ``sufficiently'' random, the generalization of the
photonic localization scenario appears applicable.

We finally comment on the (semi)classical border $n_i\to\infty$.
Increasing $n_i$ from 50 to 200, the quantum localization for
$|F_0^{\rm av}|=0.02$ almost vanishes (Figs.~\ref{fig:neffave}b and
\ref{fig:tprobave}b).  This indicates that a delocalization border is
present in the kicked atom similar to that found for the harmonically
driven system \cite{cas87}. The presence of a delocalization border
implies a maximum $n_i$ for the applicability of the photonic
localization theory.  Detailed studies of this border in the kicked
Rydberg atom remain to be performed.

\subsection{Characterization of parametric fluctuations}

\begin{figure}[h]
  \includegraphics[width=5.5cm,angle=-90]{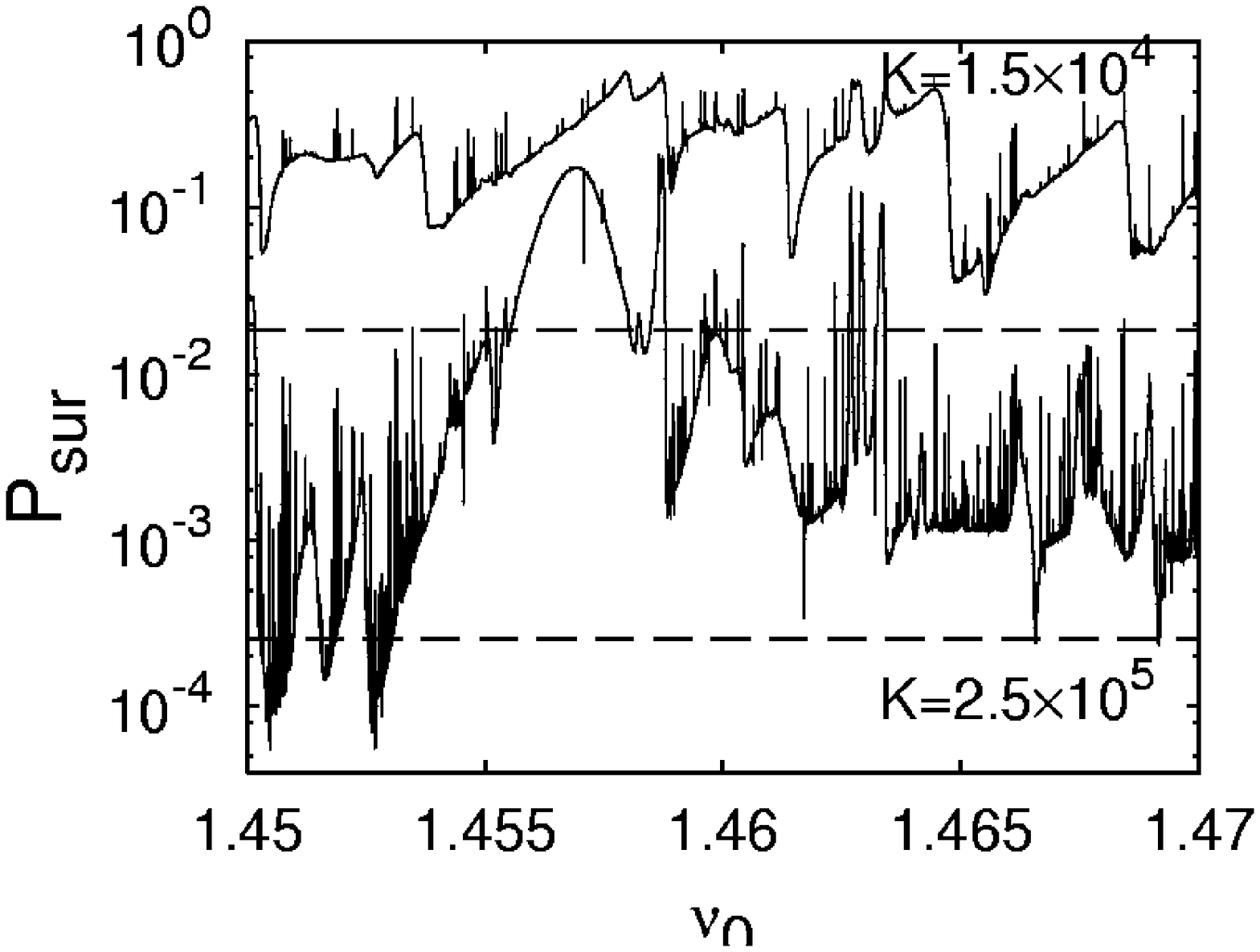}
  \includegraphics[width=5.5cm,angle=-90]{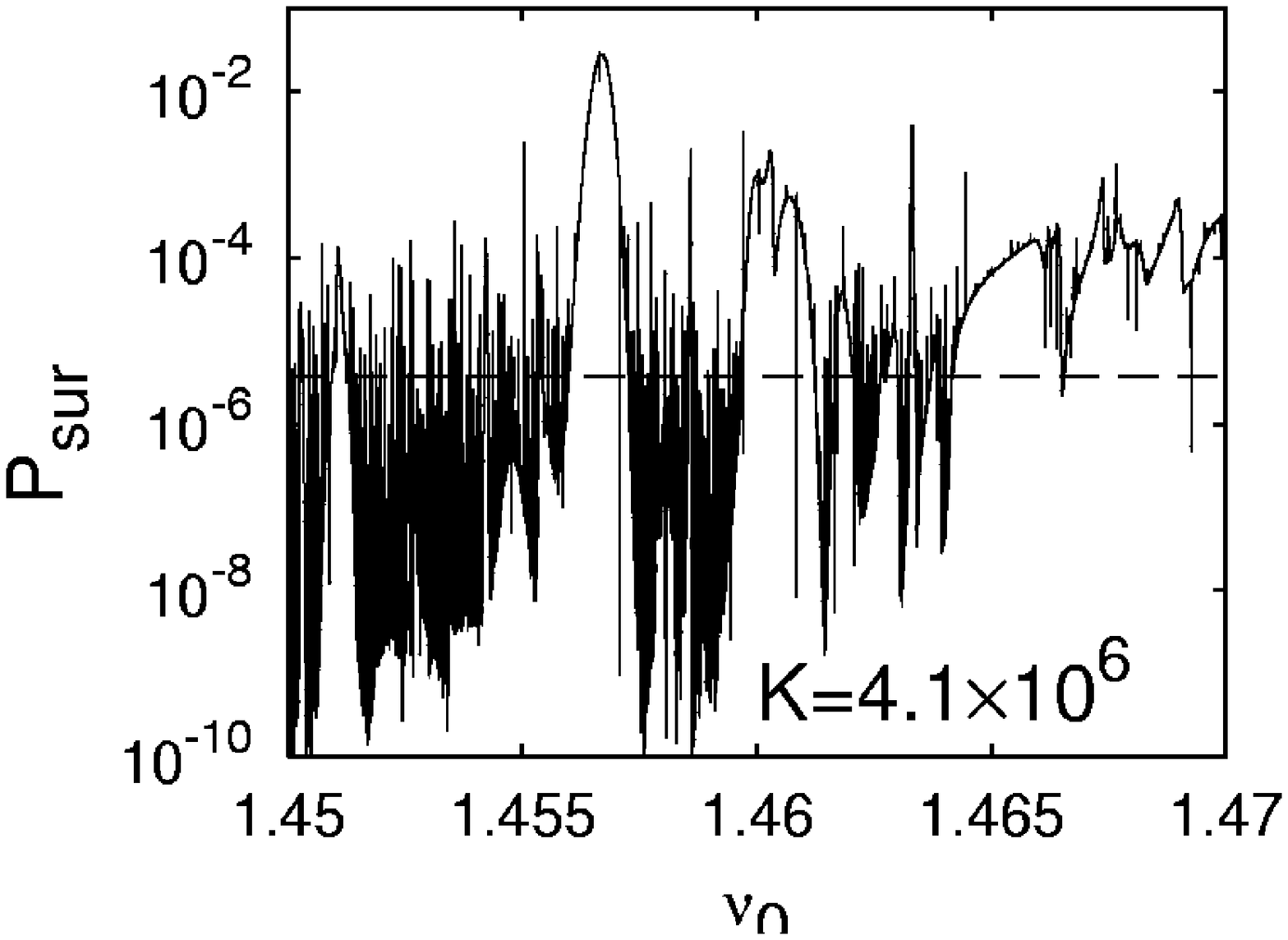}
\caption{
  Survival-probability $P_{\rm sur}$ versus frequency $\nu$ for some
  numbers $K$ of kicks. $n_i=50$ and $|F_0|^{\rm av}=0.005$.  The
  dashed lines give the classical results.  }
\label{fig:psurnu}
\end{figure}

We turn now to a quantitative description of the parametric
fluctuations of $P_{\rm sur}^{\rm qm}$ under variation of $\nu_0$ seen
in Fig. \ref{fig:psurK}.  Fig.~\ref{fig:psurnu} displays the evolution
towards an increasingly complex fluctuation pattern as $K$ increases.
The amplitudes of the fluctuations increase by several orders of
magnitude.  Strong fluctuations in the survival probability have also
been found e.g.  in the harmonically driven Rydberg atom \cite{buch98}
and in the kicked rotor when subjected to a varying Aharonov-Bohm flux
\cite{casati01} or a variation of the kicking frequency
\cite{sandro06}.

\begin{figure}[h]
  \includegraphics[width=5.5cm,angle=-90]{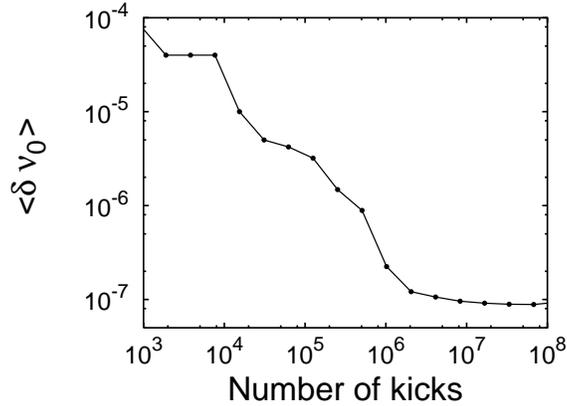}
\caption{
  Average distance $\langle\delta_{\nu_0}(K)\rangle$ between extrema
  in $P_{\rm sur}$ as a function of number $K$ of kicks.  $n_i=50$ and
  $|F_0^{\rm av}|=0.005$.  We analyze a set containing $10^6$ points
  on $1.45\le\nu_0\le 1.45008$.  }
\label{fig:avedist}
\end{figure}

In order to characterize the increasingly finer scale on which these
fluctuations occur, we determine the average distance between adjacent
maxima and minima on the frequency scale, $\langle
\delta_{\nu_0}(K)\rangle$.  At times smaller than the delocalization
time $\tau_D$, $\langle\delta_{\nu_0}(K)\rangle$ is rapidly decreasing
(Fig.~\ref{fig:avedist}).  A similar build-up of fluctuations in the
survival probability with time is observed for the kicked rotor
\cite{sandro06}.  In the very-long time-limit beyond $\tau_D$, where
the dynamics of the kicked system is governed by the decay of the
ground state, no further fluctuations are produced but
$\langle\delta_{\nu_0}(K)\rangle$ saturates.

It is now tempting to inquire whether the complexity of the
fluctuations in the $P_{\rm sur}^{\rm qm}$ can be described by an
approximate fractal dimension.  For example, the fluctuations in the
kicked rotor in the chaotic regime have been shown to have a fractal
structure \cite{casati01,sandro06}.  Fractal conductance fluctuations
have been predicted \cite{ketz} and experimentally found, e.g.
\cite{hegger}, for the transport through cavities with mixed classical
phase space. A fractal structure has also been found in the survival
probability of a system with a mixed phase space \cite{casati00}. In
both cases, a semiclassical explanation based on on a classical
power-law decay has been proposed.  Since the kicked Rydberg atom
displays classically such a power-law decay even though its phase
space is fully chaotic, rather than mixed, a fractal is a conceivable
candidate.

\begin{figure}[h]
  \includegraphics[width=5.5cm,angle=-90]{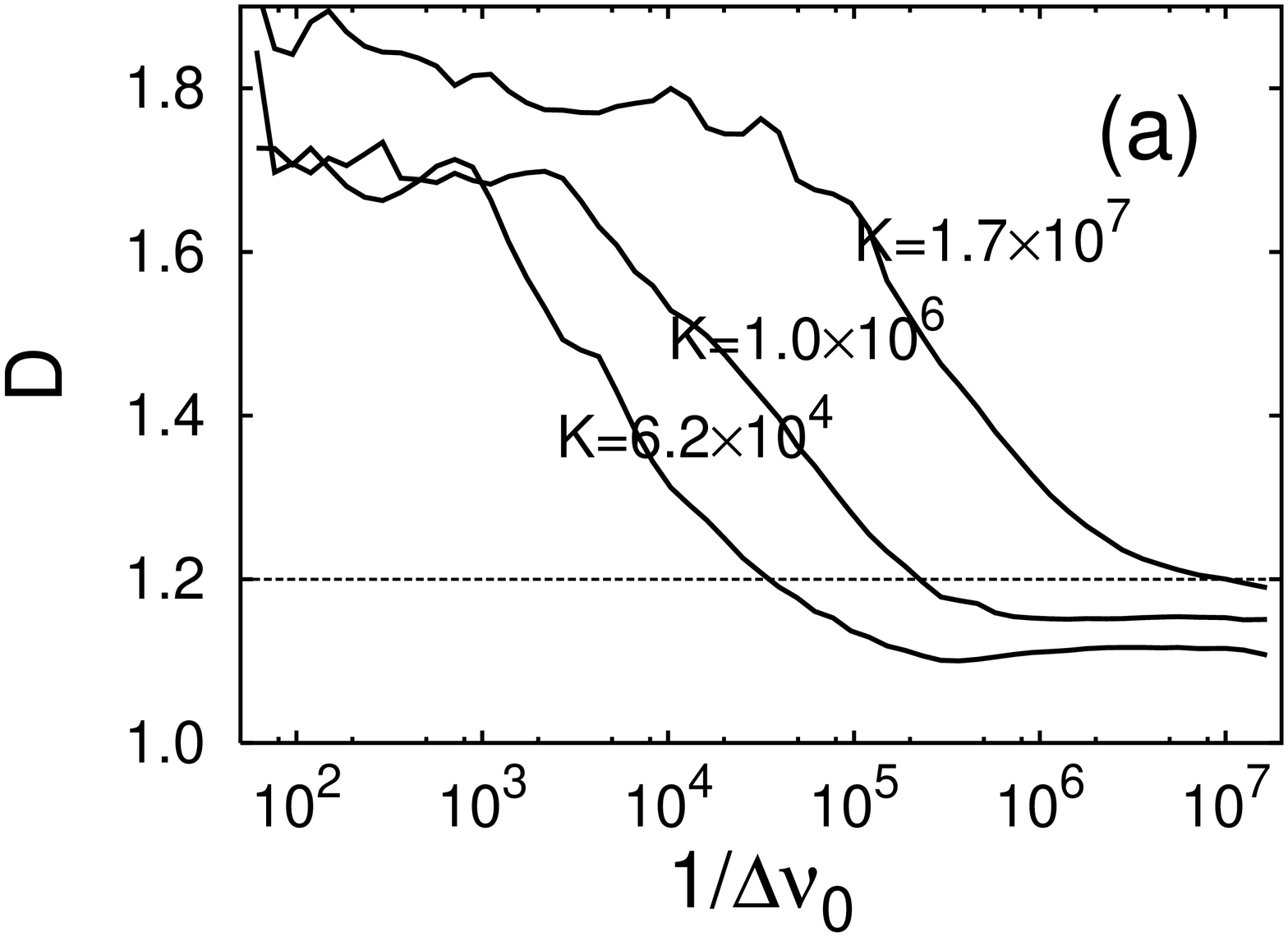}
  \includegraphics[width=5.5cm,angle=-90]{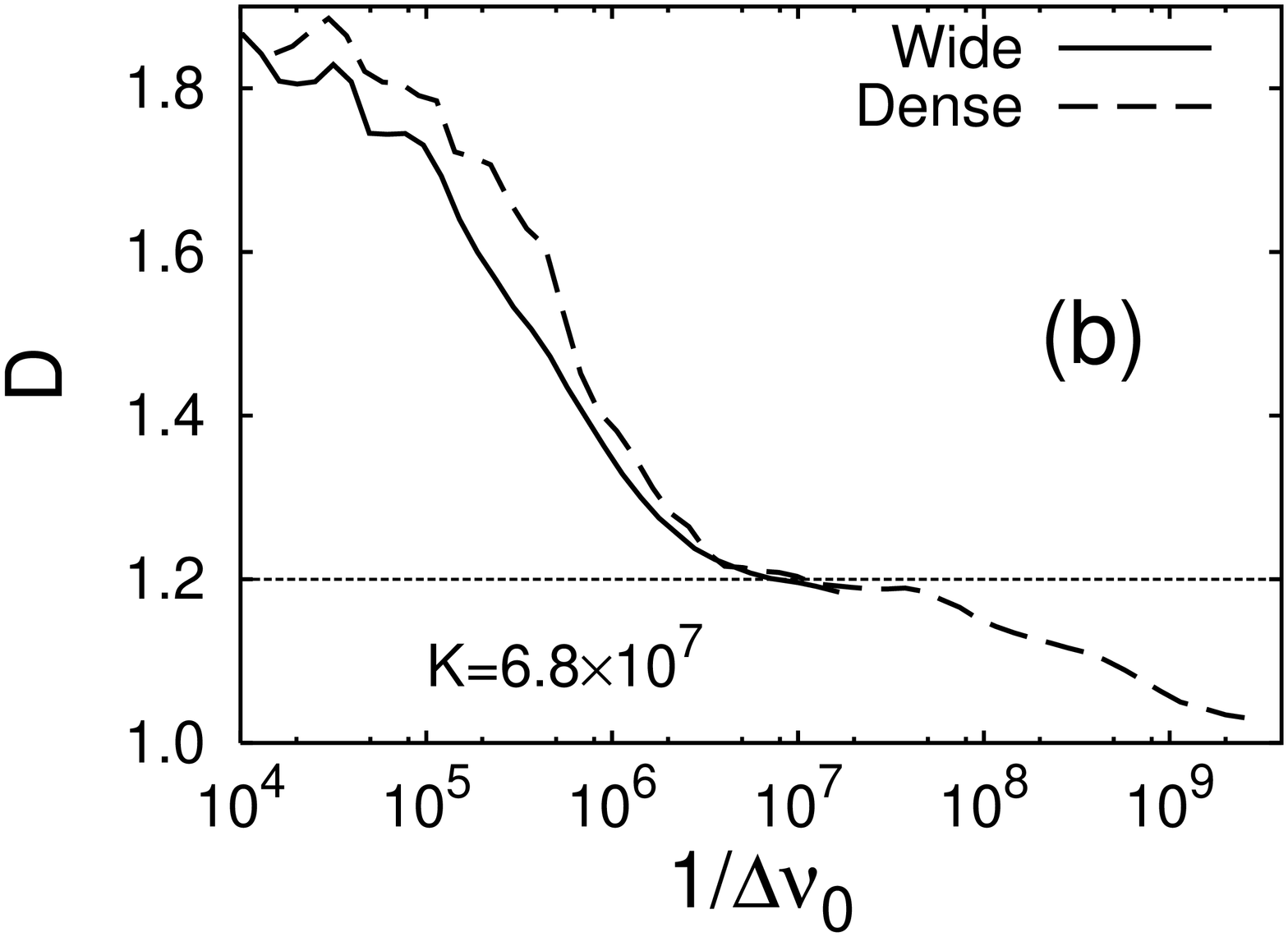}
\caption{
  Fractal analysis of $\log_{10}(P_{\rm sur})$ for $n_i=50$ and
  $|F_0^{\rm av}|=0.005$.  In {\bf (a)} we show the data analyzed for
  $2\times 10^6$ points in the interval $1.45\le\nu_0\le 1.47$ for
  different numbers $K$ of kicks and in {\bf (b)} we combine the wide
  set of (a) with data for a dense set containing $10^6$ points in the
  interval $1.45\le\nu_0\le 1.45008$ for $K=6.8\times 10^7$.  }
\label{fig:fract}
\end{figure}

For a given ``resolution'' $\Delta\nu_0$ in frequency we calculate the
local fractal dimension $D(\Delta\nu_0)$ by means of a variational
method \cite{varmeth}. Rigorously, this value should be independent of
$\Delta\nu_0$. In the present physical context, we are satisfied if
$D(\Delta\nu_0)$ is approximately constant over at least one order of
magnitude in $\Delta\nu_0$.  The fractal analysis of the data shown in
Fig.~\ref{fig:psurnu} is presented in Fig.~\ref{fig:fract}a. Since the
survival probability fluctuates over many orders of magnitude, we
analyze the logarithm of the data.  After more than $10^5$ kicks, a
plateau starts to develop, the width of which reaches almost two
orders of magnitude for larger $K$.  The value of $D$ on the plateau
increases with $K$ up to about $D\approx 1.2$.  The process of
increasingly finer rescaling is transient and stops beyond $K>10^7$
when $\tau_D$ is reached.  Zooming in on a narrow frequency interval
for $K>K_D$, (Fig.~\ref{fig:fract}b), the plateau value is still is
$D\approx 1.2$, but the width of the plateau is only about one order
of magnitude.  Thus, a tendency towards a non-integer dimension can be
found in the positively kicked Rydberg atom. The value is weakly
dependent on the number $K$ of kicks for which the fractal analysis is
made and appears to approach $D\approx 1.2$.

It is now instructing to compare this value with the semiclassical
prediction \cite{ketz,casati00,lai92} based on the power-law decay
with exponent $\alpha=1.5$ (see Fig.~\ref{fig:psurK}) of the classical
survival probability $P_{\rm sur}^{\rm cl}(K)$, $D_{\rm
  SC}=2-\alpha/2=1.25$. The value for $D_{\rm SC}$ is remarkably close
to the plateau value found from the fractal dimension analysis.  We
thus conclude that the onset of a self-similar fluctuation pattern can
be observed with a dimension close to the semiclassical prediction
derived from the classical power-law decay.  We note that this process
is transient in that the finest scale for the fluctuations is
determined by the time $\tau_D$ beyond which the most stable bound
state decays.

\section{Summary}
\label{sec:summary}

We have studied the long-time limit of quantum localization of the
positively kicked Rydberg atom, involving clear signatures of a
quantum suppression of classical ionization.  We compare the
localization as seen in the survival probability to that seen in an
average quantum number $\langle n \rangle$ describing the position of
the localized part of the wavefunction (see Eq.~(\ref{eq:neff})~).  In
$\langle n \rangle$ clearer signatures of quantum localization
prevailing to higher field strengths and larger quantum numbers are
found.  Two cross-over times could be identified. The cross-over from
classical-quantum correspondence to localization ($\tau_l$) and the
destruction or delocalization at a much later time ($\tau_D$).
Remarkably, beyond $\tau_D$, the quantum system decays faster then the
classical counterpart due to direct transitions to the continuum 
resembling photoionization. 
This process, identified in the present paper within a 1D model, 
is expected to be operative in a full 3D model as well.
Quantum localization is accompanied by strong fluctuations 
in the survival probability after a given number of kicks as a function 
of the frequency of the driving field. The average
distance between the fluctuations
$\langle\delta_{\nu_0}(t)\rangle\propto 1/t$ until $t=\tau_D$,
whereafter $\langle\delta_{\nu_0}\rangle$ saturates.  In the
localization regime, the complex fluctuation pattern approaches a 
fractal.

In this paper we have highlighted effects caused by the higher
harmonics in the driving field distinguishing the kicked atom from the
Rydberg atom driven by a microwave field. Further studies comparing
these two systems, including an assessment on how closely the quantum
localization in the kicked Rydberg atom is related to Anderson
localization using e.g. the methods in \cite{cas87,buch98}, is left
for forthcoming studies.

\begin{acknowledgements}
  This work was supported by the FWF (Austria) under grant no SFB-016.
  Discussions with Andreas Buchleitner and Shuhei Yoshida are
  gratefully acknowledged.
\end{acknowledgements}

\end{document}